\documentclass[10pt,a4paper]{emulateapj}
\usepackage{lmodern}
\usepackage[latin1]{inputenc}
\usepackage[english]{babel}
\usepackage{amsmath}
\usepackage{amssymb}
\usepackage{latexsym}
\usepackage{epsfig}
\usepackage{subfigure}
\usepackage{natbib}
\usepackage{setspace}
\usepackage{longtable}
\usepackage{graphicx}
\usepackage{adjustbox}
\usepackage{changepage}
\usepackage{rotating}
\usepackage{blindtext}
\usepackage{enumitem}
\usepackage{xcolor}
\usepackage[11pt]{moresize}
\usepackage{hyperref}
\hypersetup{backref=true, pagebackref=true, hyperindex=true, colorlinks=true,                
	breaklinks=true, urlcolor= magenta, linkcolor= red, bookmarks=true,                 
	bookmarksopen=false, filecolor=cyan, citecolor=blue, linkbordercolor=blue}

\usepackage{hyperref}
\usepackage{lineno}

\newcommand{\wfirst}{\textit{Roman}}

\shorttitle{Detecting stellar-mass black holes by \wfirst}
\shortauthors{Sajadian and Sahu}
\begin{document}

\title{Detecting isolated stellar-mass black holes by the \wfirst~telescope}

\author{Sedighe Sajadian\altaffilmark{1}}
\affiliation{Department of Physics, Isfahan University of Technology, Isfahan 84156-83111, Iran}
\altaffiltext{1}{Email: s.sajadian@iut.ac.ir}

\author{Kailash C.  Sahu\altaffilmark{2}}
\affiliation{Space Telescope Science Institute, 3700 San Martin Drive, Baltimore, MD 21218, USA\\}
\altaffiltext{2}{Email: ksahu@stsci.edu}
\affiliation{Institute for Advanced Study, Einstein Drive, Princeton, NJ 08540, USA}

\begin{abstract}
Isolated Stellar-Mass BlackHoles (ISMBHs) are potentially discernible through microlensing observations because they are expected to be long-duration microlensing events. In this work, we study detecting and characterizing ISMBHs with the \wfirst~observations. We simulate a big ensemble of these events as seen by \wfirst~and estimate the errors in the physical parameters of the lens objects, including their masses, distances, and proper motions through calculating Fisher and Covariance matrices. Since the $\sim$2.3-year time gap between \wfirst's first three observing seasons and the others may lower the efficiency of realizing microlensing events and characterizing ISMBHs, we additionally consider a scenario where we add a small amount of additional observations --one hour of observations every 10 days when the Bulge is observable during the large time gap-- which is equivalent to a total of about one additional day of observations with the \wfirst\ telescope. These extra observations increase \wfirst's efficiency for characterizing ISMBHs by $\sim 1$-$2$\% and, more importantly, improve the robustness of the results by avoiding possible degenerate solutions. By considering uniform, and power-law mass functions ($dN/dM \propto M^{-\alpha}$, $\alpha=2,~1,~0.5$) for ISMBHs in the range of $[2,~50] M_{\odot}$, we conclude that the \wfirst~telescope will determine the physical parameters of the lenses within $<5\%$ uncertainty, with efficiencies of $21\%$, and $16$-$18\%$, respectively. By considering these mass functions, we expect that the \wfirst~telescope during its mission will detect and characterize $3$-$4$, $15$-$17$ and $22$-$24$ ISMBHs through astrometric microlensing, with the relative errors for all physical parameters less than $1,~5,~10\%$, respectively. Microlensing events owing to ISMBHs with a mass $\simeq 10$-$25 M_{\odot}$ and located close to the observer with $D_{\rm l} \lesssim 0.5 D_{\rm s}$ while the source is inside the Galactic disk can be characterized with least errors.\\ 
\end{abstract}

\section{Introduction}	
A black hole (BH) refers to a massive object where the escape velocity from it exceeds the speed of light. Therefore, a BH can not reflect any light. However, it radiates what is called the Hawking radiation \citep{1974NaturHawking}, which is generally faint \citep{2022Malyshev,2022Auffinger}. 

Their formation mechanisms are as follows: (a) BHs can be formed by the death of massive stars with an initial mass higher than $20 M_{\odot}$ \citep[][]{1998ApJBailyn,2001ApJFryer,2018BHreview1}. (b) The interstellar gas at the centre of massive galaxies can directly collapse to form massive BHs \citep{2010reviewVolenteri,2013ASSLHeiman,2019NaturWise}. (c) Initial spatial fluctuations in the early universe (during the first second after the Big Bang) could potentially lead to the formation of primordial BHs as proposed by S. Hawking \citep{1971Hawking}. 

BHs are generally classified based on their mass into three categories: (i) Super-massive BHs, (ii) Intermediate-Mass BHs (IMBHs), and (iii) Stellar-Mass BHs. 

The first class---the super-massive BHs---have masses $M \geq 10^{5} M_{\odot}$. These objects can be found at the centers of massive galaxies (such as the Milky Way Galaxy, and M87), bright quasars, and Active Galactic Nuclei (AGN). These massive objects can be detected and characterized by tracking stars near massive galaxies' centre \citep[][]{2021NatRPVolonteri}.

The second class---the IMBHs---have masses in the range of $100$-$10^{5}~M_{\odot}$ and are thought to reside at centres of globular clusters \citep[][]{2017KoliopanosIMBHsreview,2020Greene_IMBHsreview}. One method to indirectly detect these objects is through gravitational waves caused by the merging of these massive objects \citep{2016PhRvLGW1,2017PhRvLGW2}. Attempts have also been made to detect IMBHs through astrometric microlensing of background stars caused by the IMBHs \citep{2016Kains,2018Kains}.

The third class---the stellar-mass BHs---form after the gravitational collapse of massive stars. These objects have masses as high as a few tens of solar mass. The number of such BHs in our galaxy has been predicted to be more than 10 million \citep{1983Shapiro,2018Lamberts}. The lowest-mass confirmed stellar-mass BHs have a mass in the range of $3$-$4.5~M_{\odot}$ \citep{2019SciTodd,2021MNRASJayasinghe}, whereas the most massive neutron stars (NSs) confirmed up to now have masses of $\lesssim 2 M_{\odot}$ \citep{2021ApJFonseca}, so there is a mass gap between confirmed NSs and stellar-mass BHs \citep[see, e.g., ][]{2022MNRASgap}. 

Stellar-mass BHs in binary systems can be detected either through transient $X$-rays emitted by the accretion of matter (from companions or close objects) onto the BHs' surface, or through observations of Doppler shifts in the spectra of stellar companions orbiting the BHs, or through both of them \citep{1972NaturCygnus}. In these systems, the Doppler shifts provide radial velocity measurements which are used to determine the dynamic masses of BHs. Up to now, more than $65$ stellar-mass BHs have been discovered in binary systems and through $X$-ray transient observations, mostly in our galaxy \footnote{\url{https://www.astro.puc.cl/BlackCAT/}} \citep{2016BlackCAT}. This method is restricted only to cases where the stellar-mass BHs are in binary systems with luminous companion objects, thus ISMBHs cannot be detected by this method. \\ 
  
A unique and powerful method for discovering ISMBHs is gravitational microlensing, which refers to a temporary enhancement in the brightness of a background star while passing behind a massive foreground object (the so-called gravitational lens) \citep{Einstein1936,Liebes1964,1964MNRASrefsdal}. In this phenomenon, the lens could be completely dark. Hence, microlensing observations can reveal the existence of dark (or faint) and massive compact objects, e.g., stellar-mass BHs, even ones located outside of our galaxy \citep{Pac1986,sajadian_M87_HST,2017SciSahu}.

The important observing issue is that the photometric light curve alone is not sufficient to calculate the physical parameters of the lens, such as its mass, distance and proper motion. However, by additionally measuring the parallax effect and astrometric shift in the source star position which is proportional to the angular Einstein radius, $\theta_{\rm E}$, a length-scale in the lensing formalism \citep[see, e.g., ][]{1995ApJwalker,1995AAHog,1995AJMiyamoto,2000ApJdominik}), the lensing degeneracy can be resolved. Instead of measuring the astrometric motion of the source star, the interferometry observations by even ground-based telescopes can resolve the lensing images. This leads to a direct measurement of $\theta_{\rm E}$, which also resolves the lensing degeneracy \citep{2019ApJDong,2020ApJZang}. Measuring finite source effects in transit, caustic-crossing and high-magnification microlensing events is another method to estimate $\theta_{\rm E}$ and resolve the lensing degeneracy \citep{2002ApJAn}.

The first unambiguous detection of an ISMBH in the Galactic disk has been reported recently based on the combined observations by the Hubble Space Telescope (\textit{HST}) and ground-based telescopes in the microlensing event OGLE-2011-BLG-0462 \citep{sahu2022}. There were some claims that this long-duration microlensing event could also be due to lower-mass objects \citep{2022ApJLam}, but recently \citet{2022Mroz} have shown that the lower mass estimates come from systematic errors and the lens mass should be $\simeq 7 M_{\odot}$. {There were other reports of microlensing events due to ISMBHs, but their lensing parameters or the nature of the lens objects were not determined uniquely \citep{2002MNRASMao,2002Bennett,2002ApJAgol,2005ApJPoindexter,2016ApJLu}.{\it The Optical Gravitational Lensing Experiment} group (OGLE) \citep{OGLE_IV, OGLE2003_1} has also found 13 long-duration microlensing events from observations in the years 2001-2009 which were due to white dwarfs, neutron stars, or black holes \citep{2016MNRASwyrzykowski}.}

In this work, we aim to study the possible detection and characterization of ISMBHs by the \wfirst~mission. The {\it Nancy Grace Roman Telescope} will observe the Galactic-bulge field during six $62$-day seasons in its $5$-year mission \citep{2019ApJSwfirst}. Even though its observing strategy is aimed at detecting free-floating planets and exoplanets beyond the snow line, we expect that the \wfirst~telescope will also detect microlensing events due to other lens objects \citep{sajadian_habitable,2021MNRAS_sajadian}. Additionally, because of high photometric accuracy during microlensing observations, it can resolve some second-order perturbations \citep{bagheri,sajadian_disk_roman}. \wfirst\ is also expected to detect ISMBHs through observations of long-duration microlensing events. The relatively long lifespan of the \wfirst~mission is very appropriate for detecting long-duration microlensing events and measuring both annual parallax effects and astrometric trajectories of source stars. \\

The scheme of the paper is as follows. In Section \ref{simul}, we explain all the details for simulating astrometric microlensing events as seen by the \wfirst~telescope. In Section \ref{stat}, we first explain how to calculate Fisher and Covariance matrices for photometry and astrometry measurements by \wfirst~from microlensing events due to ISMBHs. Then, we illustrate the results of our simulations and statistical calculations. Finally, in Section \ref{conclusion}, we briefly review our results and conclusions.\\
\begin{figure*}
\includegraphics[width=0.49\textwidth]{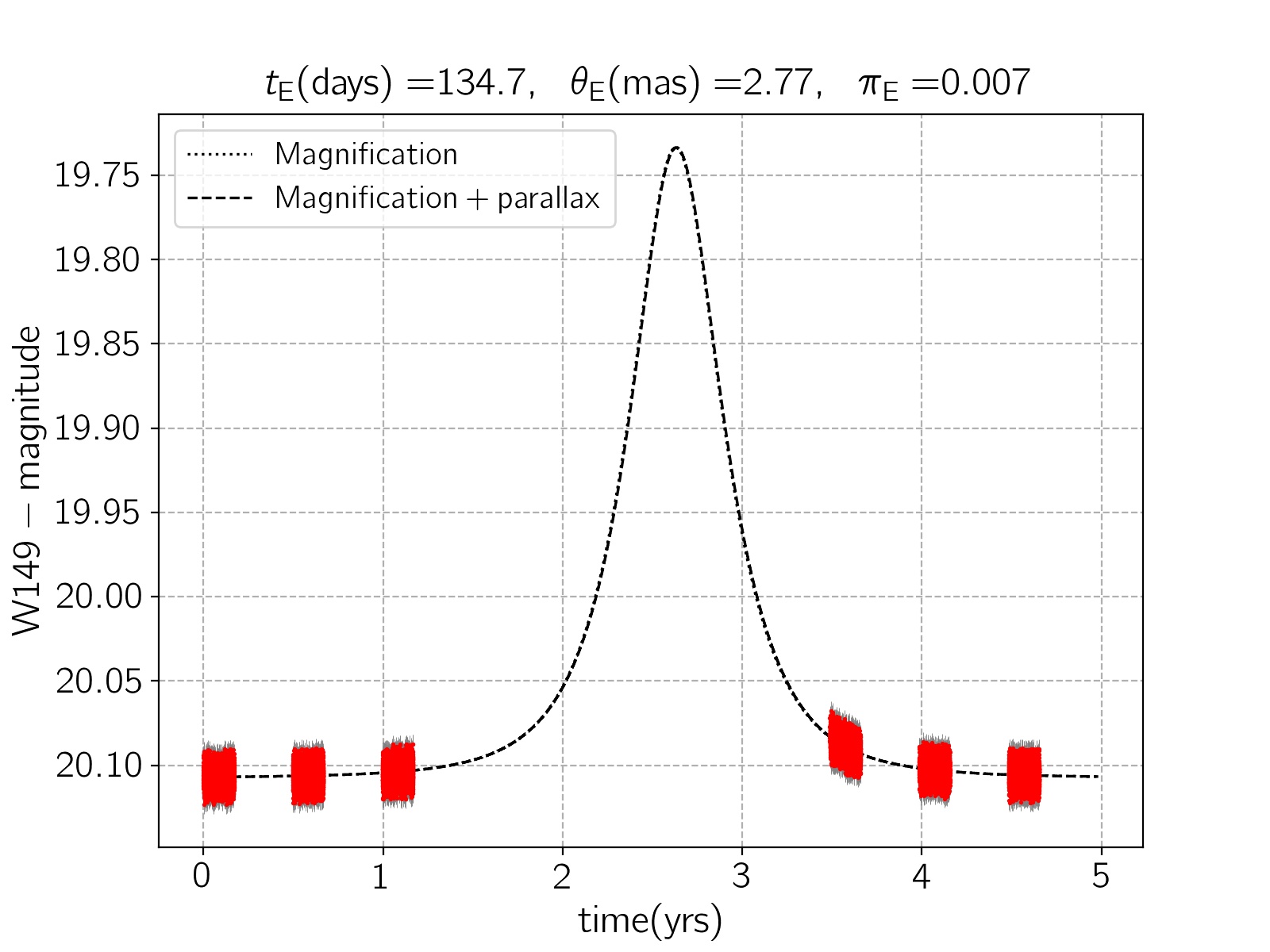}
\includegraphics[width=0.49\textwidth]{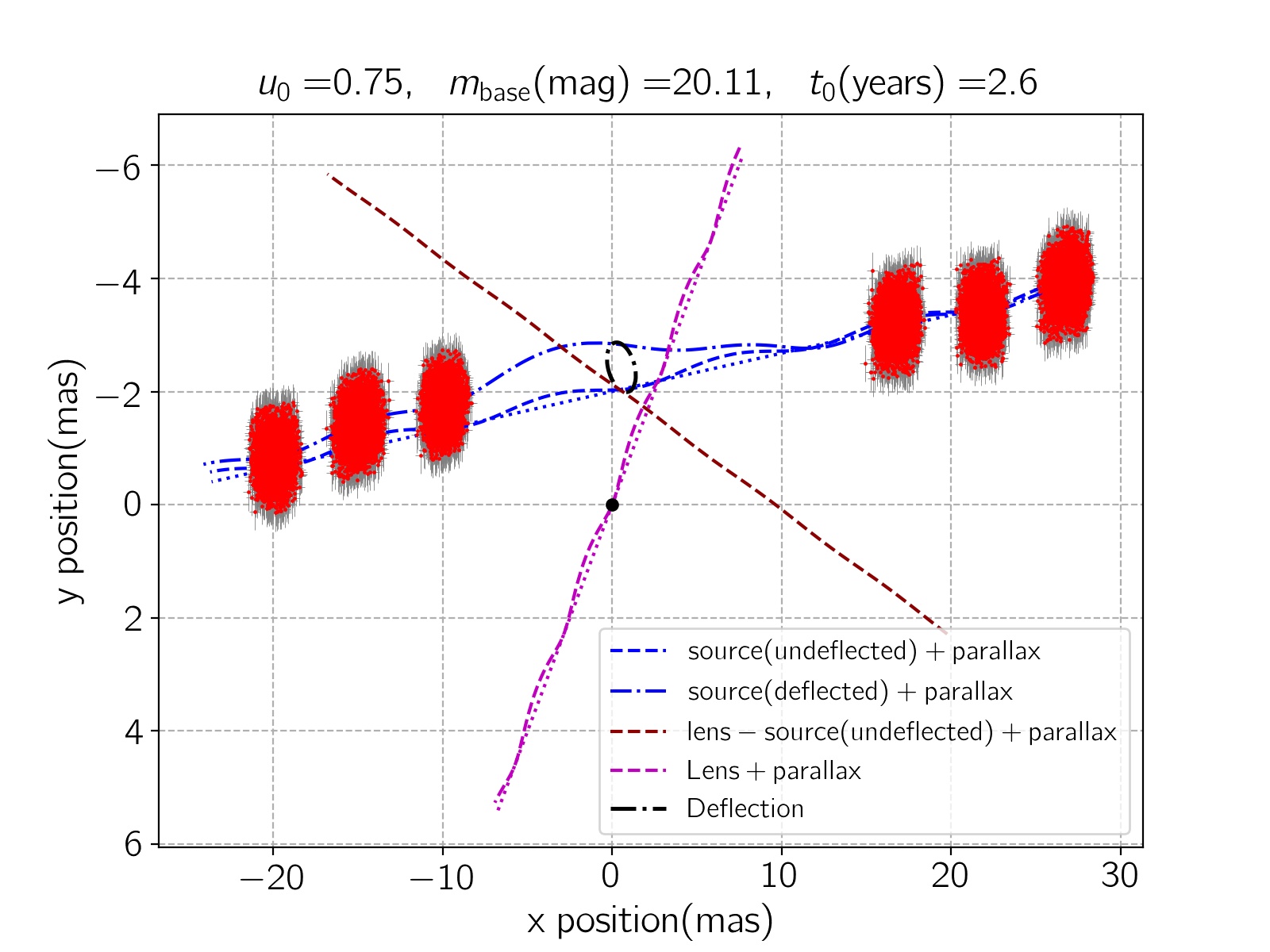}
\includegraphics[width=0.49\textwidth]{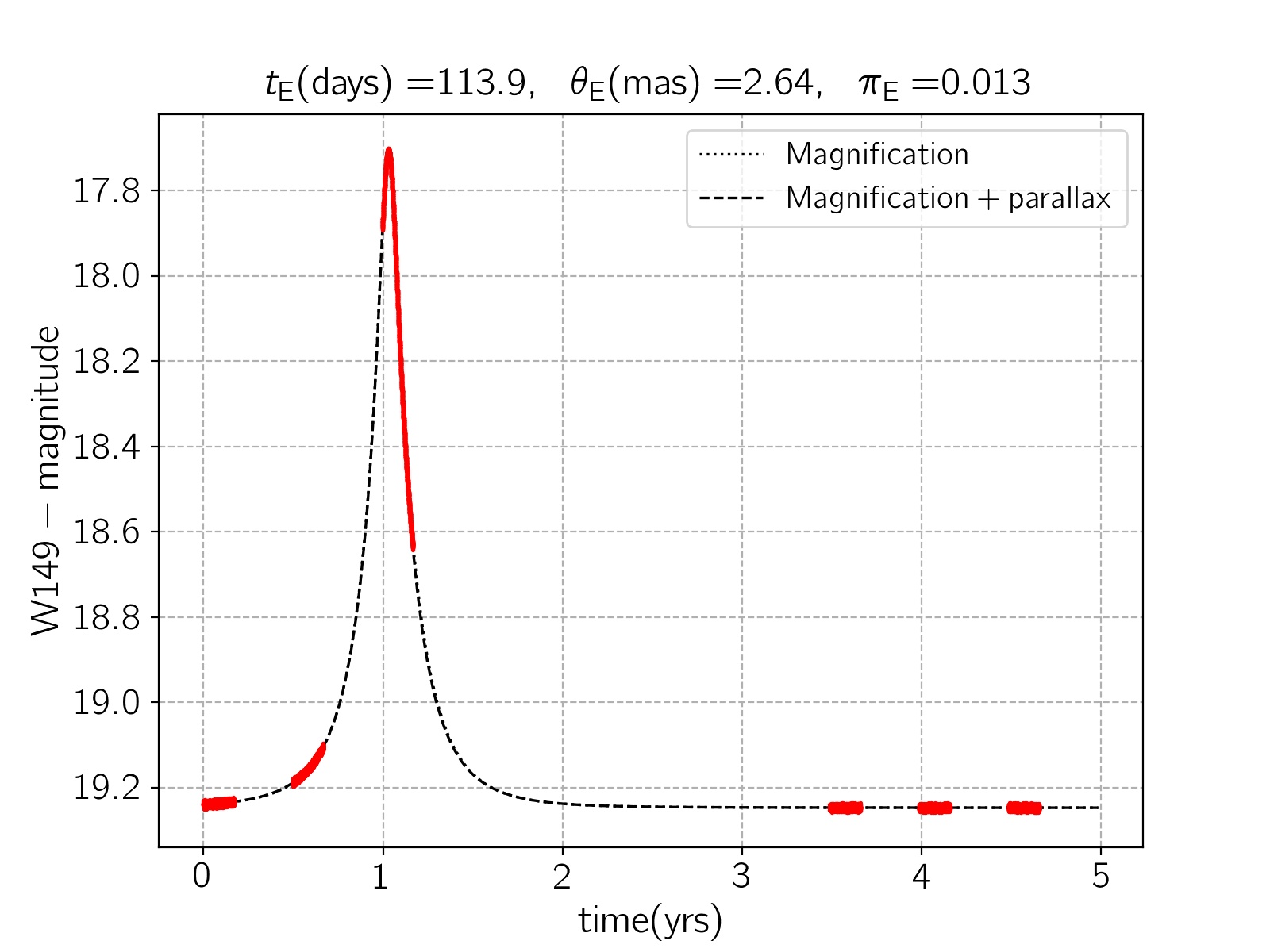}
\includegraphics[width=0.49\textwidth]{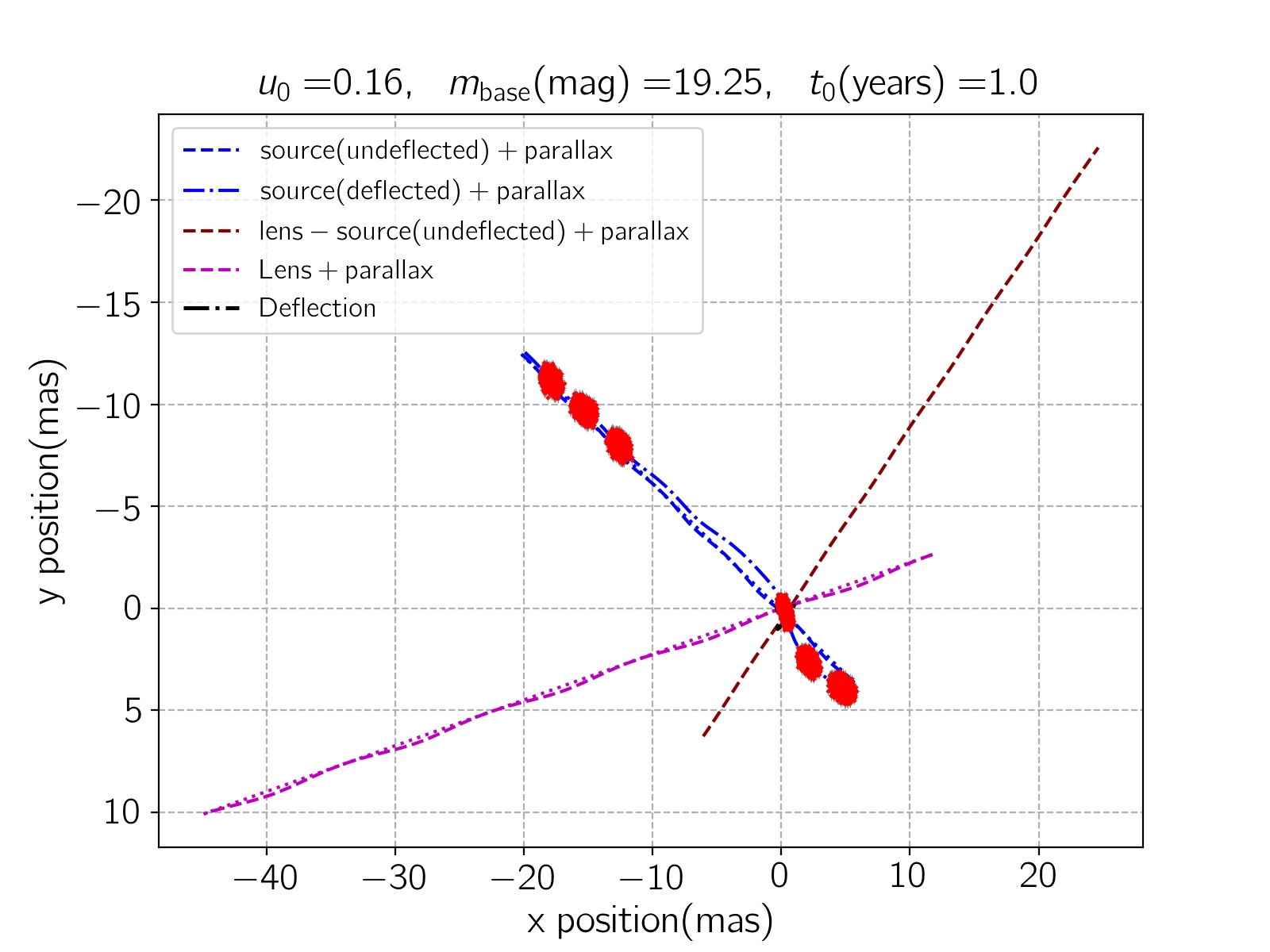}
\caption{Two examples of simulated magnification curves.
The left panels show the magnification curves with (dashed curves) and without (dotted curves) the parallax effect. 
The right panels show the corresponding astrometric motions of the source stars (blue curves), lens objects (magenta curves), and their relative motions (dark red curves)
projected on the sky plane. The synthetic data are taken with the \wfirst~telescope. The observable parameters used to make them are mentioned at the top of their lightcurves {and astrometric plots}.}
\label{exam1}
\end{figure*}

\begin{figure*}
\includegraphics[width=0.49\textwidth]{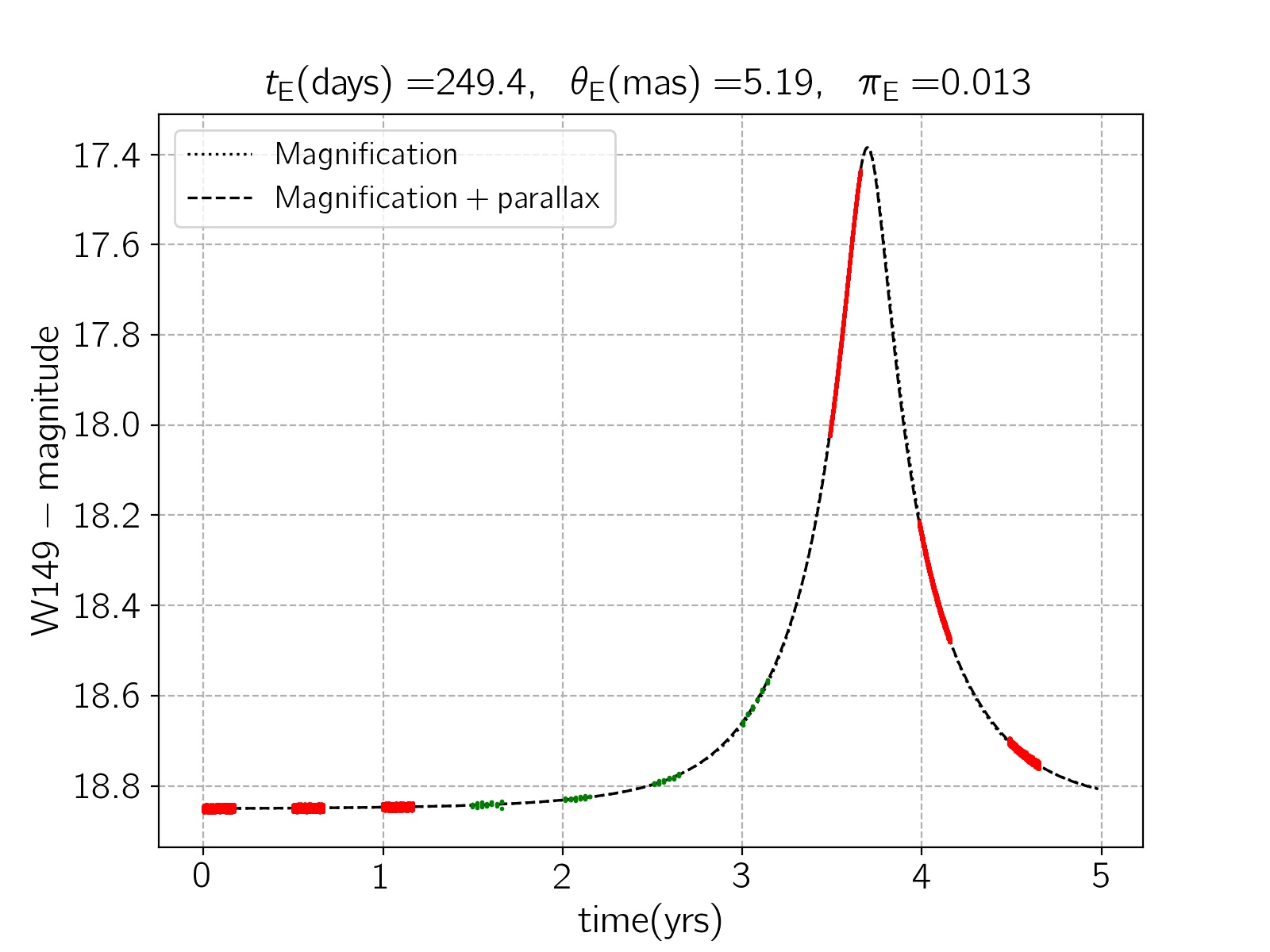}
\includegraphics[width=0.49\textwidth]{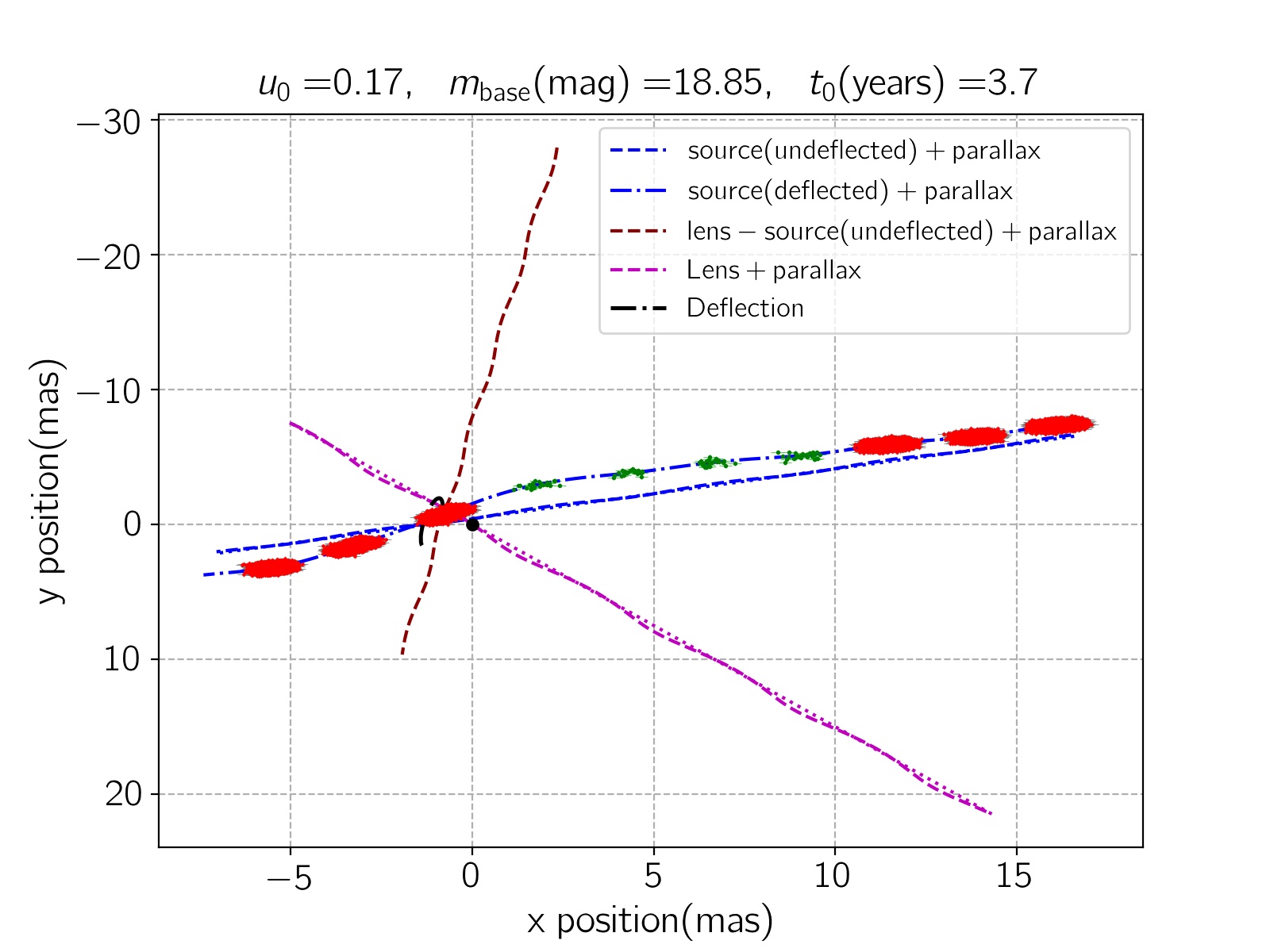}
\includegraphics[width=0.49\textwidth]{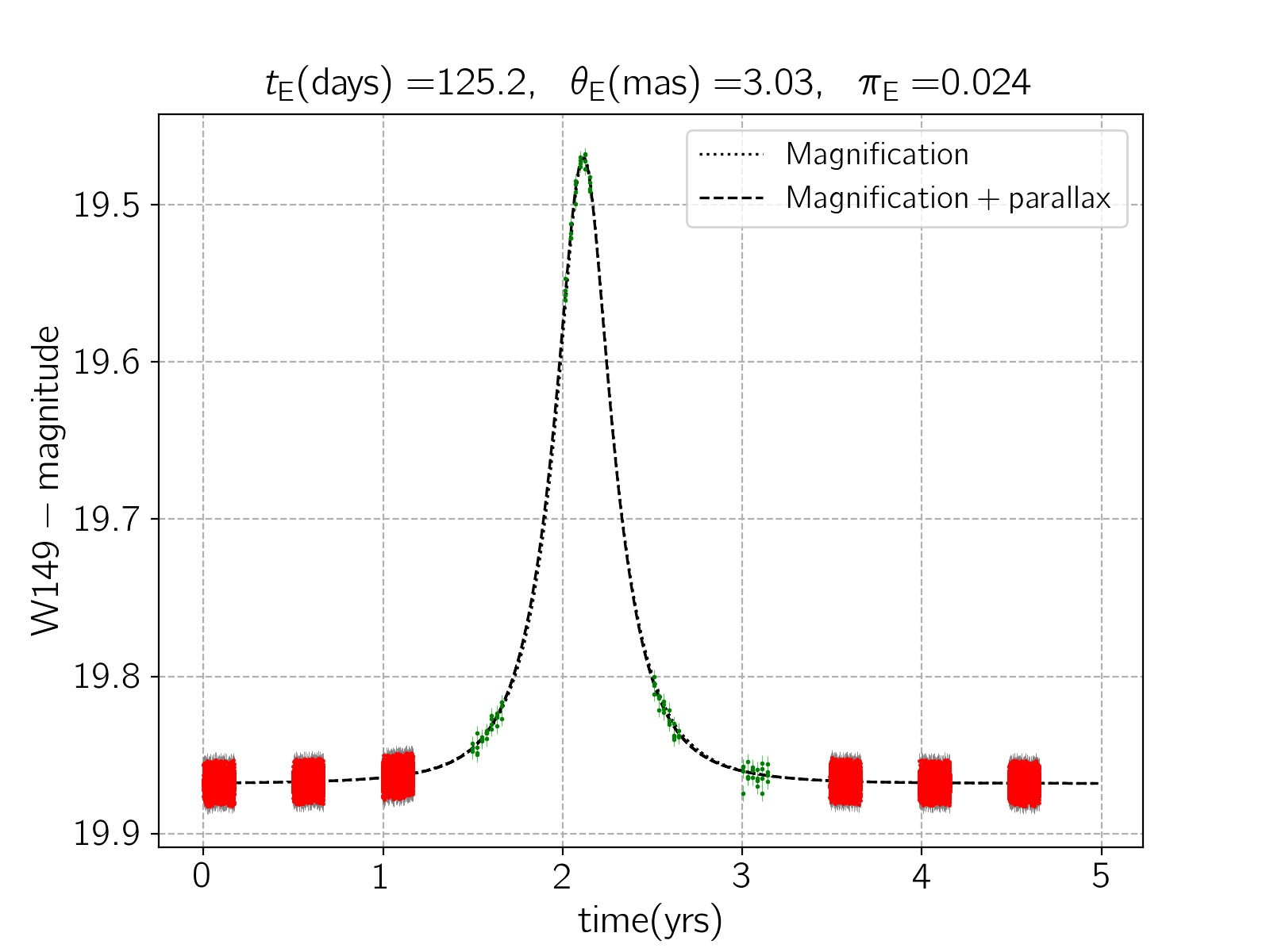}
\includegraphics[width=0.49\textwidth]{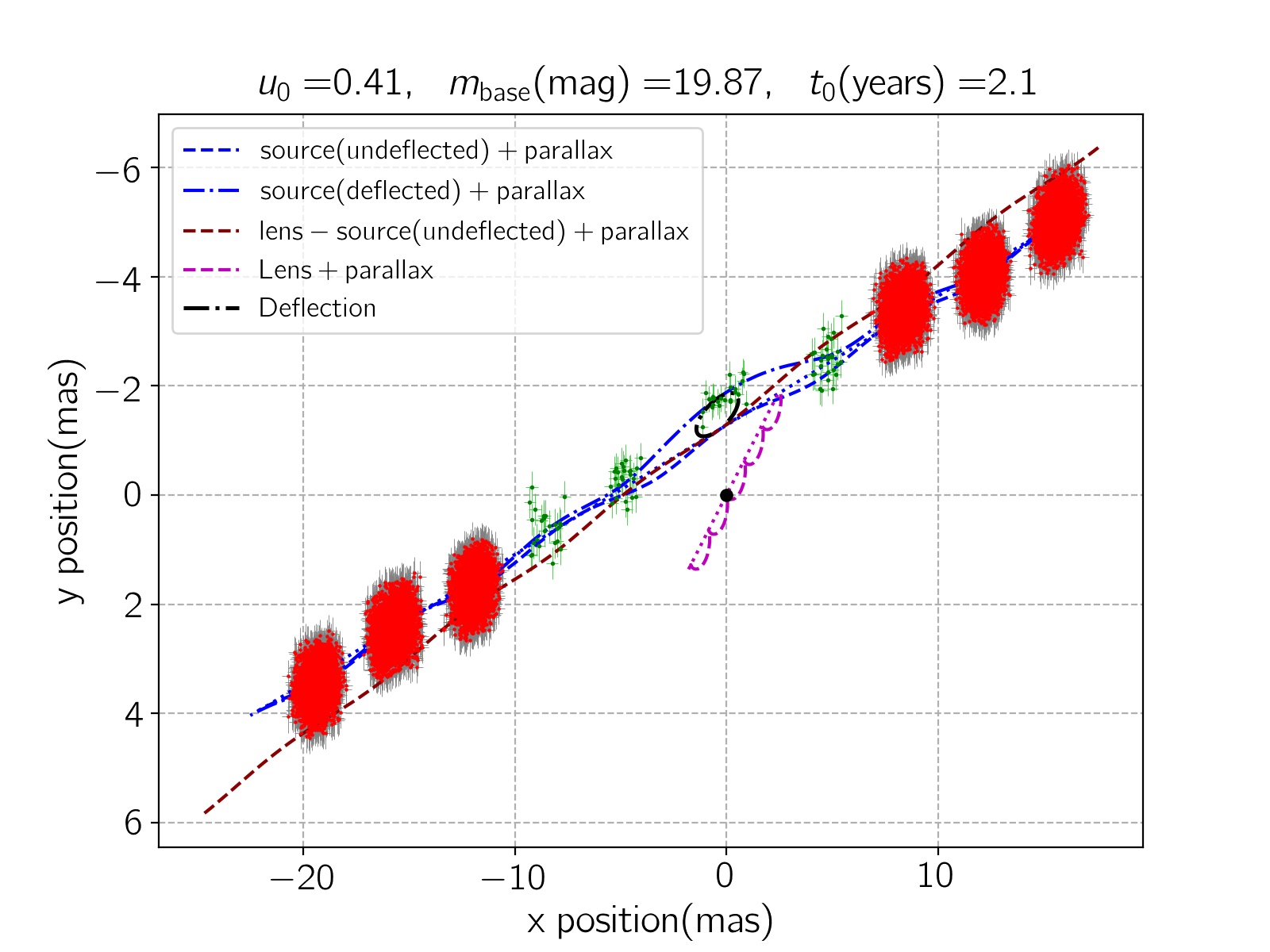}
\includegraphics[width=0.49\textwidth]{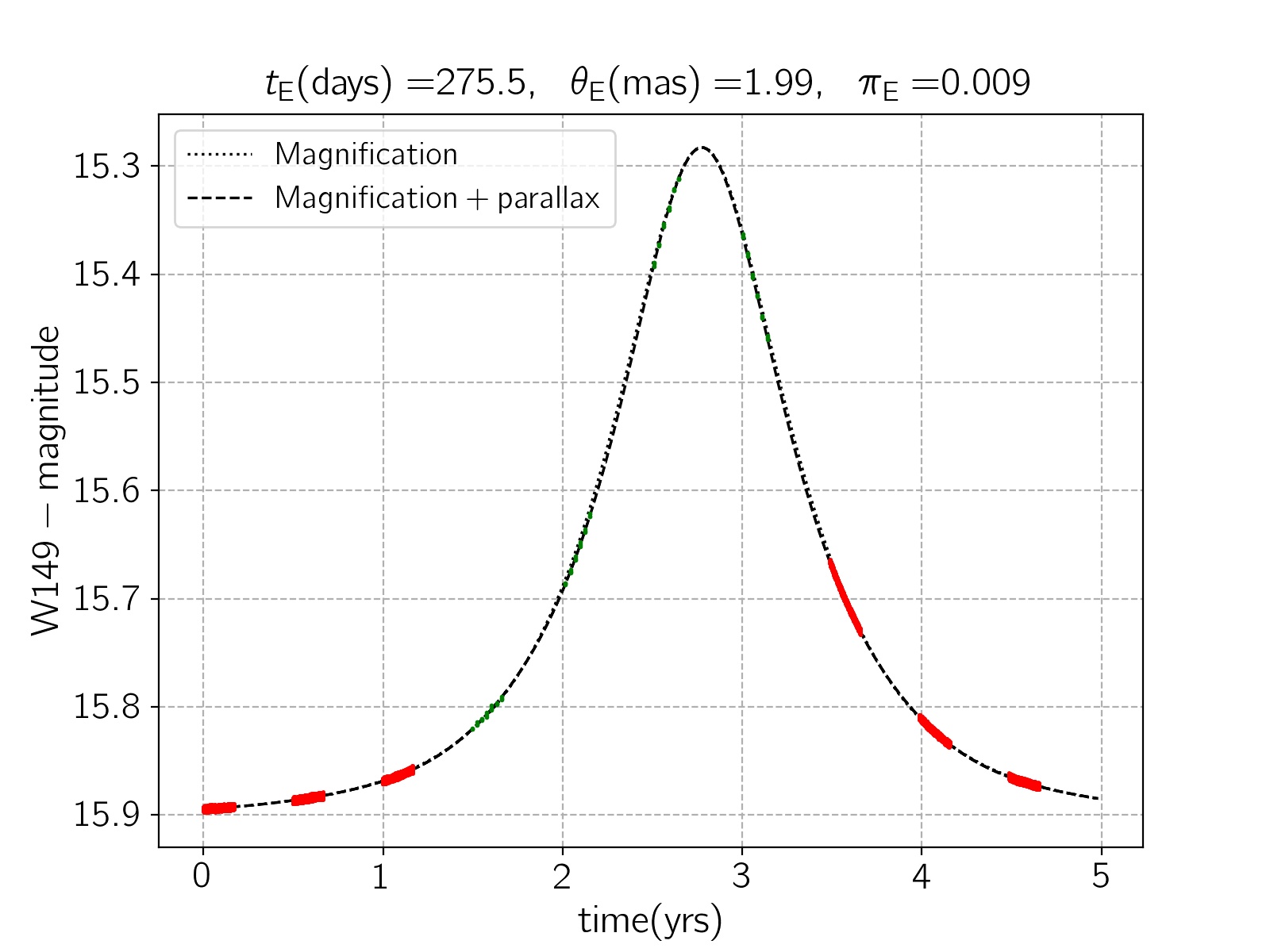}
\includegraphics[width=0.49\textwidth]{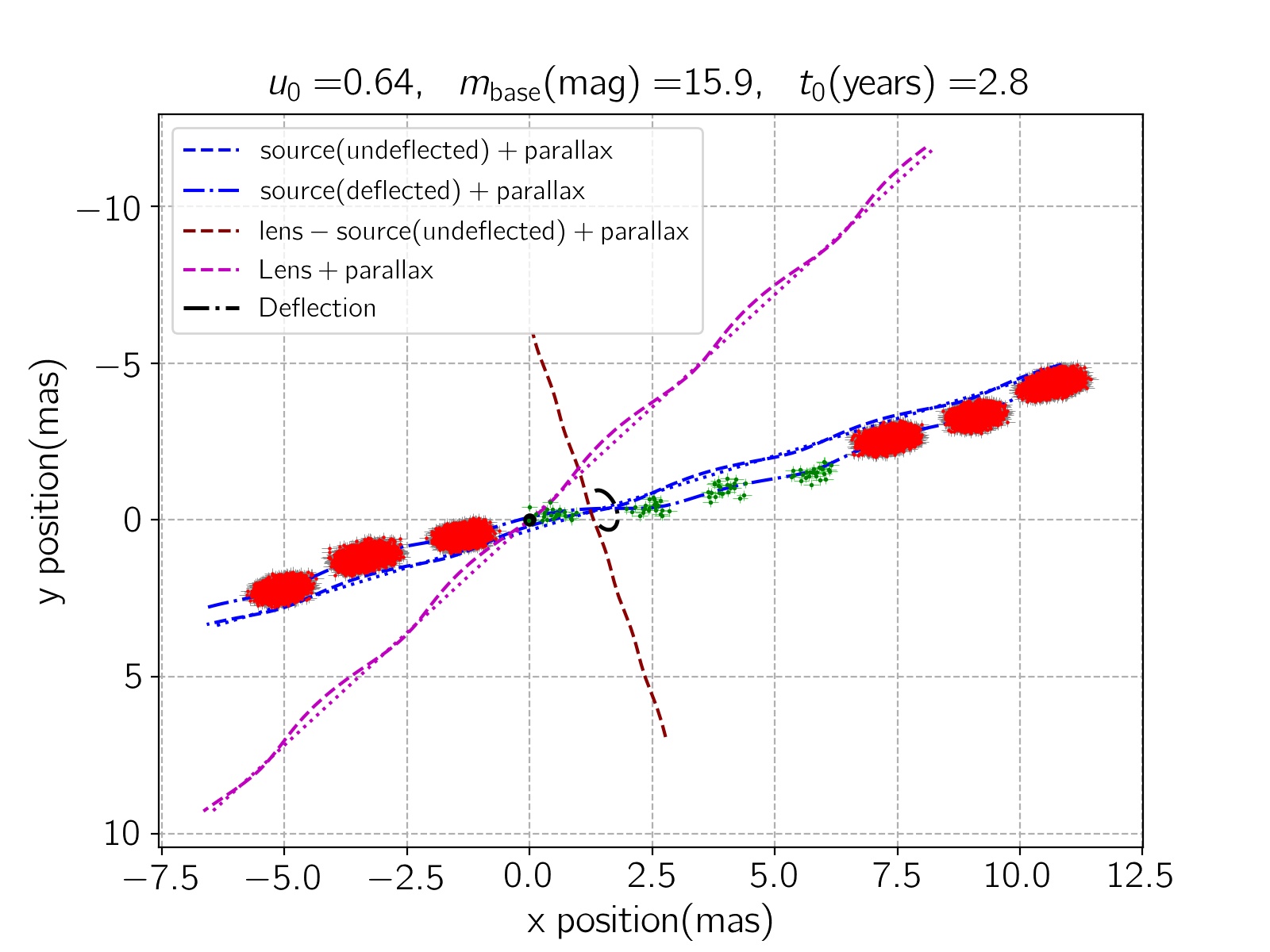}
\caption{Same as Figure \ref{exam1}, but by considering extra observations, one-hour observations of the Galactic bulge every 10 days when the Bulge is observable during the $\sim$2.3-year time gap, with the \wfirst\ telescope. These extra data points are depicted in green color.}
\label{exam2}
\end{figure*}

\section{formalism}\label{simul}
Here we review the known formalism for astrometric microlensing. We start with ignoring the parallax effect but add this at a later stage. The temporary enhancement in the stellar brightness due to the gravitational lensing of a point-like and massive object which is called the magnification factor versus time, $t$, is given by \citep[see, e.g., ][]{gaudi2012,2018Tsapras}:  
\begin{eqnarray}
A(t)=\frac{u^{2}+2}{u \sqrt{u^2+4}}, ~~~ u=\sqrt{u_{0}^{2}+ \big(\frac{t-t_{0}}{t_{\rm E}}\big)^{2}}, 
\label{magn}
\end{eqnarray}

\noindent where, $u$ is the lens-source distance projected on the sky plane and normalized to the Einstein radius (i.e., $R_{\rm E}$ the radius of the image ring at the complete alignment), $u_{0}$ is the lens impact parameter (the smallest lens-source distance), and $t_{0}$ is the time of the closest approach. The Einstein crossing time, $t_{\rm E}$, represents the lensing timescale which is given by:
  
\begin{eqnarray}
t_{\rm E}=\frac{\theta_{\rm E}}{\mu_{\rm{rel}, \odot}}=  \frac{1}{\mu_{\rm{rel}, \odot}} \sqrt{M_{\rm l}~\pi_{\rm{rel}}~\kappa},  
\label{tee}
\end{eqnarray} 
Here, $M_{\rm l}$ is the lens mass, $\kappa=8.14~\rm{mas}.M_{\odot}^{-1}$ is a constant, and $\pi_{\rm{rel}}=\rm{au}\big(1/D_{\rm l}-1/D_{\rm s}\big)$ is the relative parallax amplitude, and $D_{\rm l}$, $D_{\rm s}$ are the lens and source distances from the observer. We note that $\theta_{\rm E}=R_{\rm E}\big/D_{\rm l}$ is an angular length-scale in the lensing formalism.

\noindent $\mu_{\rm{rel}, \odot}$ is the size of the relative lens-source angular velocity. If we ignore the observer's motion around the Sun, the relative velocity vector (with respect to the Sun) is given by:  
\begin{eqnarray}
\boldsymbol{\mu}_{\rm{rel}, \odot}=  \boldsymbol{\mu}_{\rm s}- \boldsymbol{\mu}_{\rm l}= \frac{\boldsymbol{v}_{\rm s}- \boldsymbol{v}_{\odot}}{D_{\rm s}} -\frac{\boldsymbol{v}_{\rm l}-\boldsymbol{v}_{\odot}}{D_{\rm l}},  
\end{eqnarray}
where, $ \boldsymbol{v}_{\rm s}$, $ \boldsymbol{v}_{\rm l}$, and $ \boldsymbol{v}_{\odot}$ are the source, lens and the Sun velocity vectors projected on the sky plane. In Appendix \ref{append1}, we explain how to convert the stellar velocities from the Galactic coordinate frame to the observer frame. \\

{\bf Parallax effect:} We know that the observer (here, the \wfirst~telescope) rotates around the Sun, so the real relative lens-source angular velocity will be a function of time and is given by:  
\begin{eqnarray}
\boldsymbol{\mu}_{\rm{rel}}(t)= \boldsymbol{\mu}_{\rm{rel}, \odot} + \frac{\pi_{\rm{rel}}}{\rm{au}}~\boldsymbol{v}_{\rm o}(t), 
\end{eqnarray}

\noindent $\boldsymbol{v}_{\rm o}$ being the velocity vector of the observer with respect to the Sun projected on the sky plane as explained in Appendix \ref{append1} \footnote{For projection of the observer orbit on the sky plane, first we should project the observer orbit on the Galactic plane by a rotation $60^{\circ}$ around the intersection line of the orbital plane and the Galactic plane.}. Hence, the observer's rotation around the Sun, which is a function of time, causes the relative lens-source angular velocity to be a function of time, and as a result, it makes a periodic perturbation in the magnification curve, the so-called parallax effect \citep[][]{Gould1994}. By considering this effect in the lensing formalism, the normalized source-lens angular displacement (which determines the magnification factor) versus time is given by: 

\begin{eqnarray}
\boldsymbol{u}=u_{0}\begin{pmatrix}
-\sin \xi  \\ \cos \xi
\end{pmatrix} +\frac{t-t_{0}}{t_{\rm E}}\begin{pmatrix}
\cos \xi\\
\sin \xi
\end{pmatrix} + \frac{\pi_{\rm E}}{\rm{au}}\int_{t_{0}}^{t} dt\begin{pmatrix}
v_{\rm o,n1}\\
v_{\rm o, n2}
\end{pmatrix}
\label{tEE}
\end{eqnarray}

\noindent where, $\pi_{\rm E}=\pi_{\rm{rel}}/\theta_{\rm E}$  which is a dimensionless parameter, and  $\xi$ is the angle between the relative source-lens trajectory and the direction of increasing Galactic longitude, i.e. $\boldsymbol{n1}$ (as defined in Appendix \ref{append1}) which is given by $\tan \xi= \mu_{\rm rel, \odot, n2}/\mu_{\rm rel, \odot, n1}$. \\

\noindent According to the literature, we could define $\pi_{\rm E}$ as a vector which is parallel with the relative lens-source proper motion, i.e., 
\begin{eqnarray}
\boldsymbol{\pi}_{\rm E}=\big(\pi_{\rm n1},~\pi_{\rm n2}\big)= \pi_{\rm E} \big( \cos \xi,~\sin \xi \big).
\end{eqnarray}

The initial parameters that can be derived from the simple form of microlensing lightcurves (Eq. \ref{magn}) are $t_{0}$,~$u_{0}$, and $t_{\rm E}$ . In observations toward the Galactic bulge, most of the source stars are located in the Galactic bulge, at a distance $D_{\rm s}=8$ kpc from us. Measuring $t_{\rm E}$ gives us only a relation between the lens mass, the lens distance, and the relative lens-source angular velocity, even by fixing the source distance. However, discerning the parallax effect in the lightcurve allows us to measure the vector of the parallax amplitude, $\boldsymbol{\pi}_{\rm E}$, which is still not enough to resolve the lensing degeneracy completely.

{\bf Astrometric microlensing:} One way to resolve this degeneracy and determine these parameters specially for long-duration microlensing events due to ISMBHs is resolving the source angular trajectory projected on the sky plane: 
\begin{eqnarray}
\boldsymbol{\theta}_{\rm s}(t)= \boldsymbol{\theta}_{\rm s, 0}(t) + \frac{\boldsymbol{u}}{u^{2}+2}  \theta_{\rm E},   
\label{thetas}
\end{eqnarray}
\\
\noindent where, the last term is the astrometric shift in the apparent brightness center of the source star which is another result of the lensing effect. In the lensing formalism where a background star is lensed by a point-like and massive lens object, two distorted images are formed whose brightness center does not coincide with the source center. We note that this astrometric shift is proportional to the Einstein angular radius which is a function of the lens mass and its distance \citep[see, e.g.,][]{1995AJMiyamoto,2000ApJdominik}.   

In Equation \ref{thetas}, $\boldsymbol{\theta}_{\rm s, 0}(t)$, is the position vector of the source star projected on the sky plane as a function of time as seen by the observer, which is: 
\begin{eqnarray}
\boldsymbol{\theta}_{\rm s, 0}(t)=\boldsymbol{\theta}_{\rm s, 0}(t_{0})+ \boldsymbol{\mu}_{\rm s}(t-t_{0}) - \frac{1}{D_{\rm s}}  \int_{t_{0}}^{t}\boldsymbol{v}_{\rm o}(t) dt,
\label{tet0} 
\end{eqnarray}

\noindent where, the first term, $\boldsymbol{\theta}_{\rm s, 0}(t_{0})= u_{0}~\theta_{\rm E}\big(-\sin \xi,~\cos \xi\big)$, is the source position on the sky plane at the time of the closest approach with respect to the lens position (i.e., the coordinate center). The second term specifies a straight line over the sky plane. The last term, which is related to the effect of the observer's motion around the Sun on the source position, is mostly very small because of the large source distance from the observer. {This can be clearly seen by comparing the blue dotted lines (which do not take the parallax effect into account) and the blue dashed lines (which take the parallax effect into account) in the right panels of Figures \ref{exam1} and \ref{exam2}}. This term makes a periodic perturbation on the source trajectory projected on the sky plane. 

The lens also has a similar angular trajectory projected on the sky plane, given by
\begin{eqnarray}
\boldsymbol{\theta}_{\rm l}(t) = \boldsymbol{\mu}_{\rm l}(t-t_{0}) - \frac{1}{D_{\rm l}}  \int_{t_{0}}^{t}\boldsymbol{v}_{\rm o}(t) dt. 
\label{thetal}
\end{eqnarray}

\noindent Here, we have set the lens location at the coordinate center at the time of the closest approach. However, in most of the gravitational microlensing events the lens objects are dark and their angular trajectories cannot be determined. We note that 
$$\boldsymbol{u}(t)= \frac{\boldsymbol{\theta}_{\rm s}(t)- \boldsymbol{\theta}_{\rm l}(t)}{\theta_{\rm E}}$$

Let's come back to Equation \ref{thetas}, which describes the source angular trajectory projected on the sky plane versus time. In the case of astrometric observations where we discern this source trajectory, the observables that we can measure are: (a) $\theta_{\rm E}$, which is the angular size of the Einstein ring radius, (b) $\boldsymbol{\mu}_{\rm s}$,  the angular source velocity projected on the sky plane with respect to the observer, and (c) the sign of the lens impact parameter \citep[e.g.,][]{2015MNRASs}.

\noindent However, for discerning the second one, observations are necessary either long after or long before the lensing event. Additionally, the astrometric shift due to lensing effect has longer timescale than $t_{\rm E}$. It tends to zero as $u^{-1}$, while the magnification factor is proportional to $\propto u^{-4}$ for $u\gg1$ \citep[see, e.g., ][]{2000ApJdominik}. Its long timescale helps to resolve the time dependent perturbations, such as the orbital-motion effect in binary lensing \citep{sajadian_orbital}. \\

By measuring both astrometric shift due to microlensing and the parallax effect in the magnification curve, we determine $t_{\rm E}$, $\theta_{\rm E}$, $\pi_{\rm E}$, $\xi$, and $\boldsymbol{\mu}_{\rm s}$, which allows us to completely resolve the lensing degeneracy and determine $D_{\rm l}$, $M_{\rm l}$, $\boldsymbol{\mu}_{\rm{rel}, \odot}$, and $\boldsymbol{\mu}_{\rm l}$. We note that $u_{0}$, and $t_{0}$ are measurable from magnification curve and are necessary while modeling the astrometric motion of the source star, but they are not directly involved in extracting the physical parameters.\\

One class of microlensing events that are specially interesting are the long-duration events caused by ISMBHs. In these events, the astrometric shift in the source angular position is considerable, because of the large angular Einstein radius. Additionally the parallax effect potentially could be measured, because of long duration of such events. We note that in most of the microlensing events due to ISMBHs, the finite source effect is negligible, unless the lens passes over the source surface. This is is rare since the impact parameter has to be less than the normalized angular source radius, $u_{0}<\rho_{\rm s}$, $\rho_{\rm s}= \theta_{\rm s}/\theta_{\rm E}$, where $\theta_{\rm s}$ is the angular source radius, and the large value of  $\theta_{\rm E}$ decreases $\rho_{\rm s}$.

\begin{figure*}
\includegraphics[width=0.49\textwidth]{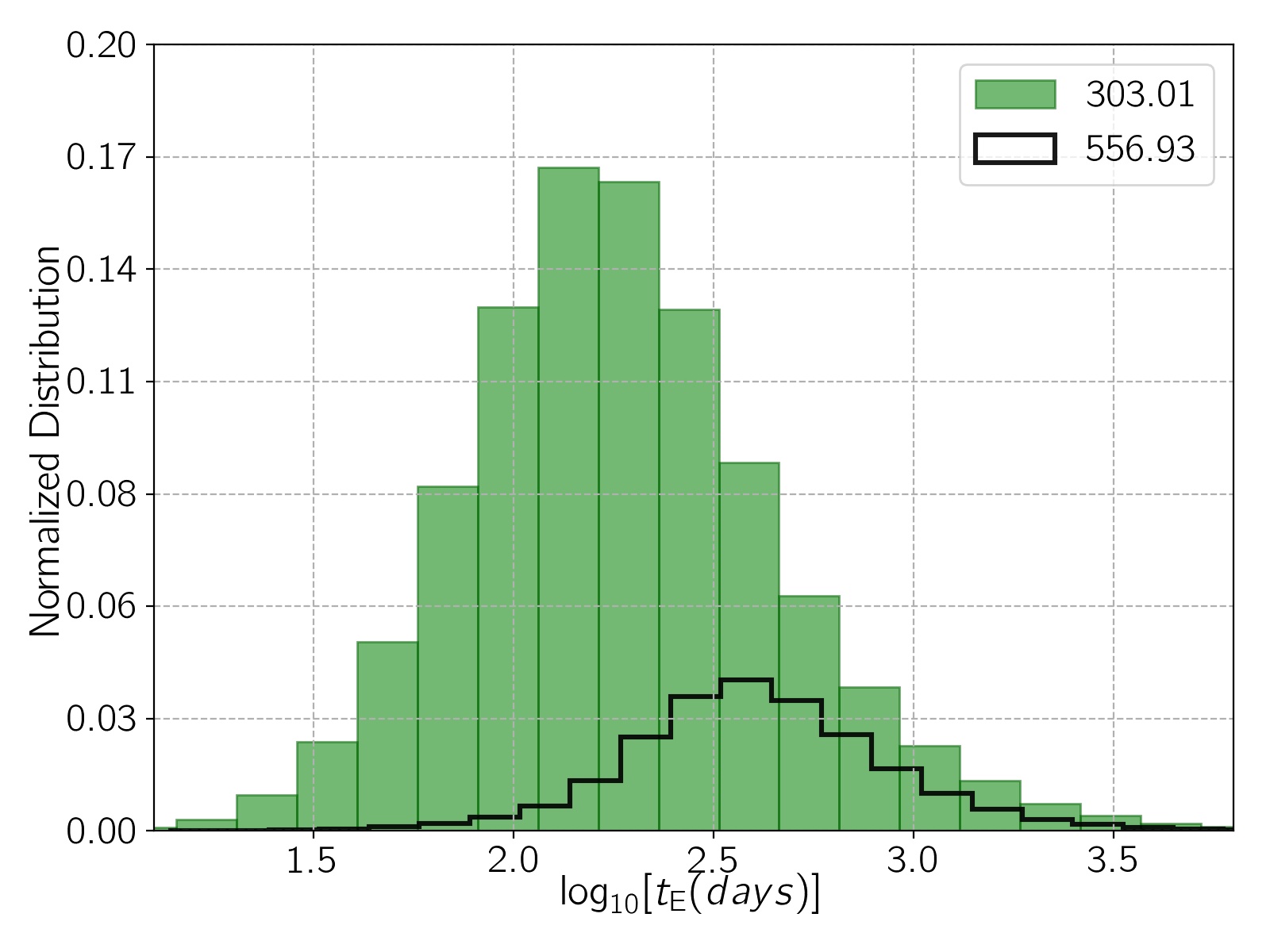}
\includegraphics[width=0.49\textwidth]{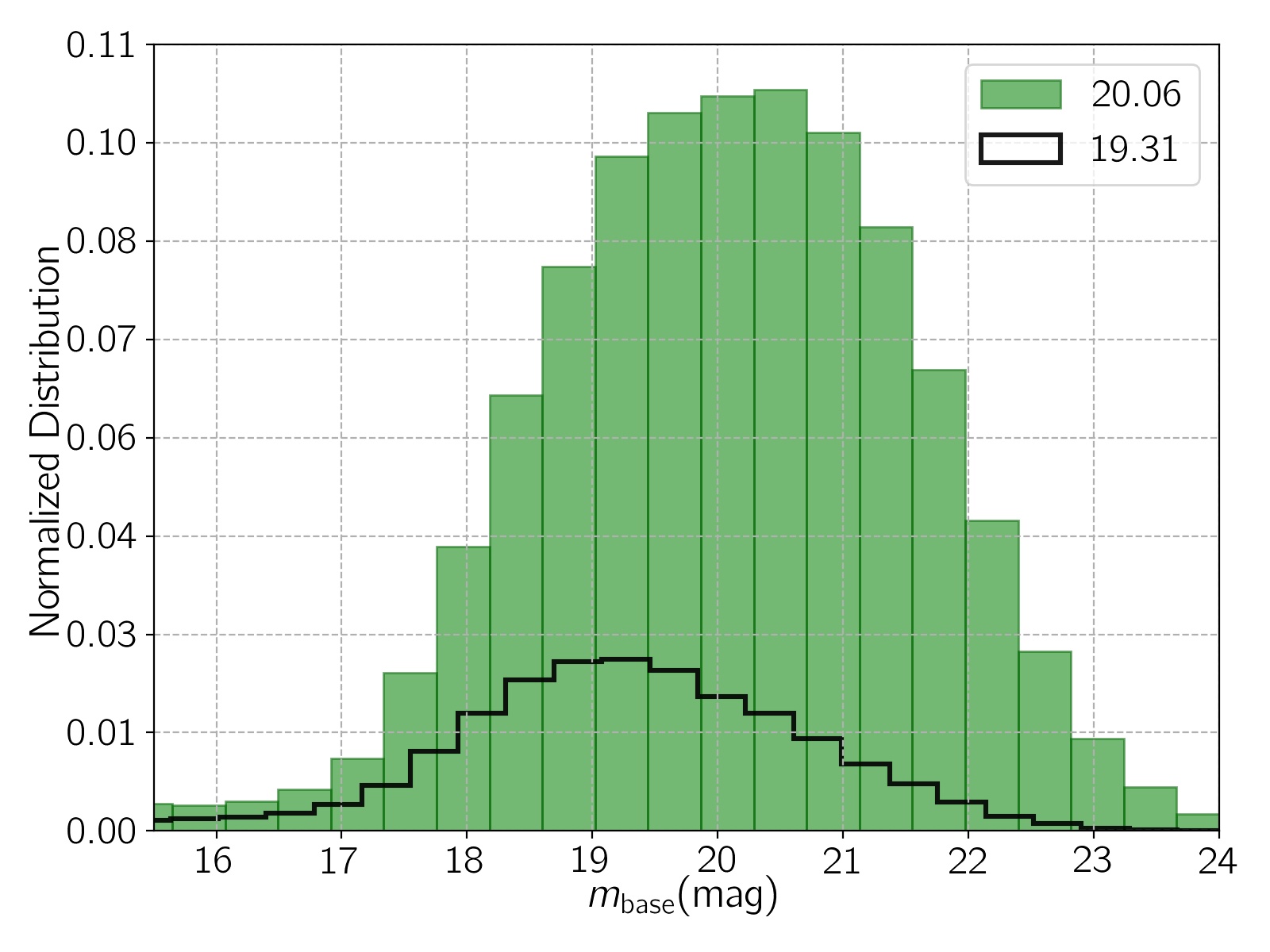}
\includegraphics[width=0.49\textwidth]{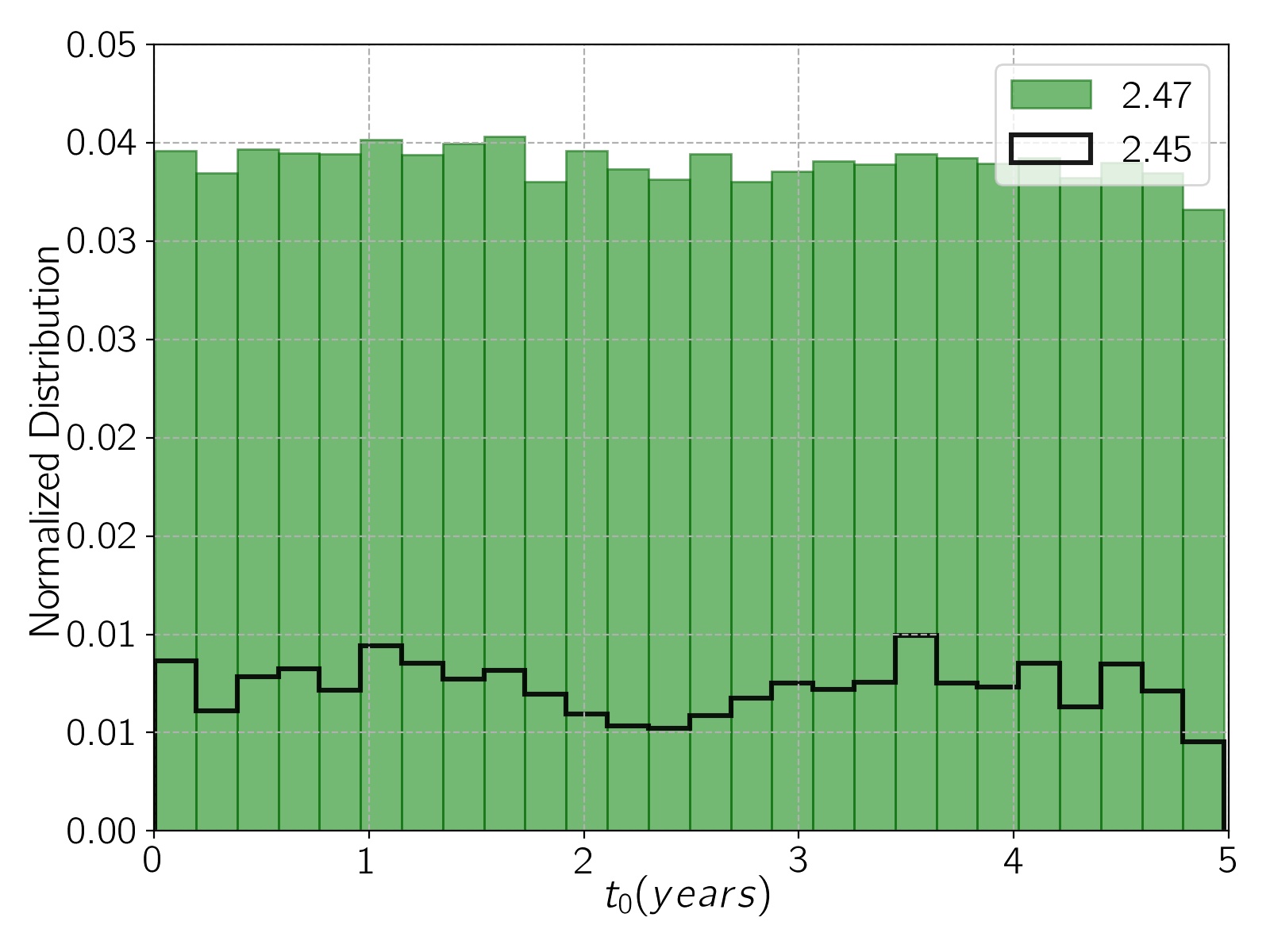}
\includegraphics[width=0.49\textwidth]{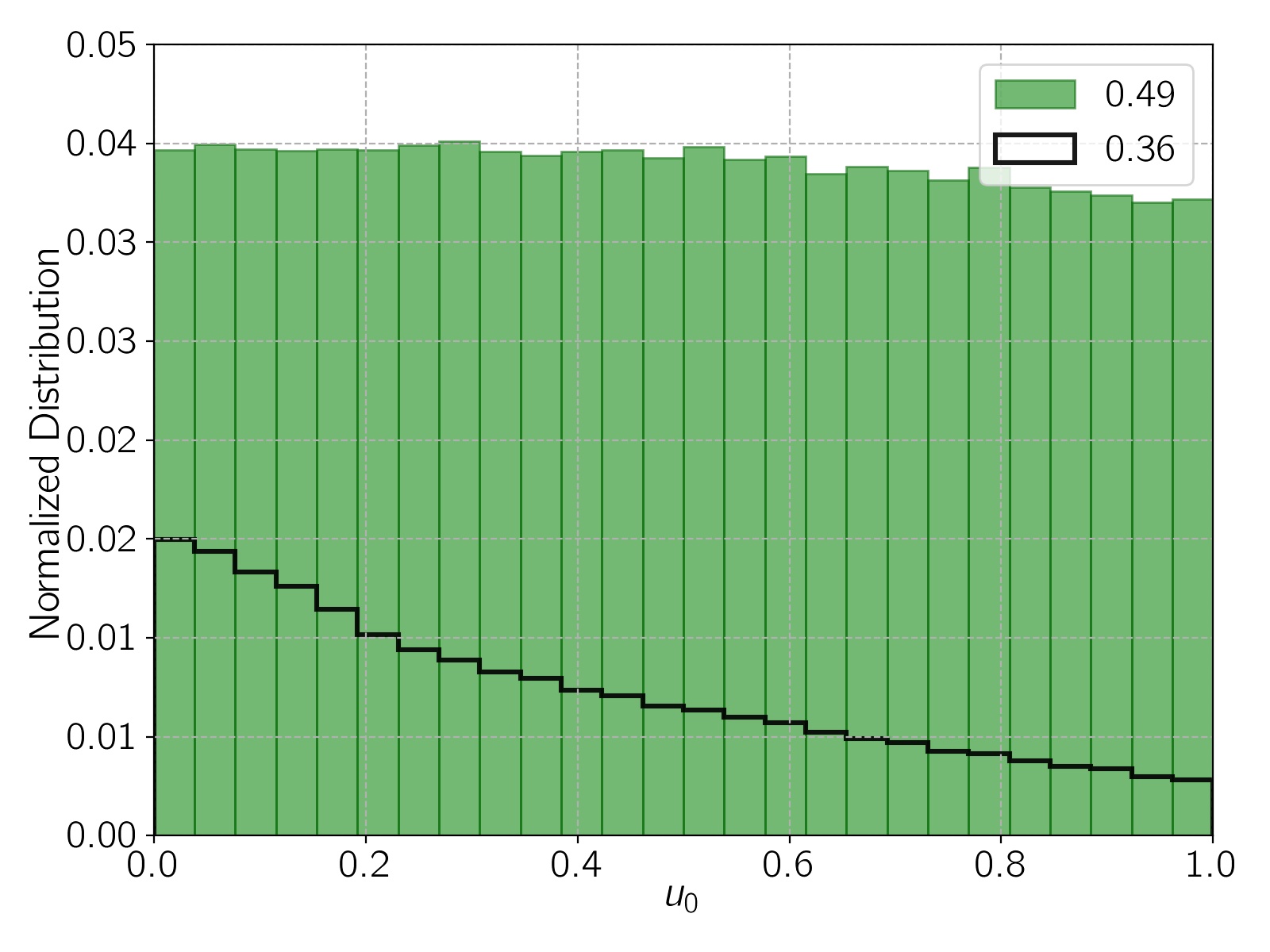}
\caption{The normalized (fractional) distributions of $t_{\rm E}$, $m_{\rm{base}}$, $t_{0}$, and $u_{0}$ for all the detected microlensing events by \wfirst\ are depicted in green. Also, the normalized distributions of the events for which the physical parameters of the lenses are measurable with $\leq 5\%$ relative errors  (after considering the extra observations during $\sim$2.3-year time gap) are shown as black stepped curves. The average values of these parameters calculated from related distributions are mentioned in the legends.}
\label{hist}
\end{figure*}

Using the introduced formalism, we simulate the astrometric microlensing events due to ISMBHs toward the Galactic bulge. We also make the synthetic data points according to the \wfirst~observing strategy. In this regard, the observing cadence is fixed at $15.16$~min. The observations include six $62$-day seasons, three of them at the first part of the \wfirst~mission with a time interval $110$-day between seasons, and three other seasons at the end.

The photometric observations are mostly done in the W149 filter. This filter roughly corresponds to $W149=(K+H+J)\big/3$ \citep{2017Montet}. Its photometric precision, $\sigma_{\rm m}$, is a function of the apparent magnitude \citep{2019ApJSwfirst,2020Johnson}. The astrometric precision of the \wfirst~observations also strongly depends on the apparent stellar brightness. S. Calchi Novati (private communication) has modelled the \wfirst\ astrometric precisions for stars of different magnitudes through \textit{Jitter} simulations and in this work we use his simulations to determine the \wfirst\ astrometric precision. He has used the \wfirst\ observing strategy described by \citet{2019ApJSwfirst}, and calculated the astrometry precisions through simulations \citep[see, e.g.,][]{astrometric_kepler}.\\

Two examples of simulated astrometric microlensing events are shown in Figure \ref{exam1}. The left panels show the magnification curves with (dashed curves) and without (dotted curves) the parallax effect and their corresponding right panels show the related astrometric motions of the source stars (blue curves), lens objects (magenta curves), and their relative motions (dark red curves). The observable parameters which characterize these events are specified at the top of the light curve and astrometric motion plots. 

There is a large time gap of $\sim$2.3 years between the first three and the last three observing seasons of \wfirst \footnote{\url{https://roman.gsfc.nasa.gov/galactic_bulge_time_domain_survey.html}}, which lowers the detection efficiency of ISMBHs. If the peak of the light curve happens during this large time gap (which lasts $\sim 2.3$ years), discerning such events will have large uncertainties, and several degenerate models will fit the data well. For instance, the peak of the first lightcurve in the top panel of Figure \ref{exam1} was not covered by \wfirst\ data which would have been useful in correctly determining the microlensing parameters, including the parallax. 
	
\noindent Hence, for a robust determination of the microlensing parameters, we additionally consider a case where the \wfirst\ telescope observes the seven Galactic-bulge fields for a total of one hour every 10 days when the Galactic bulge is observable during the 
$\sim$2.3-year time gap. Although these observations are sparse and use a total of $\sim$1-day of \wfirst\ time, they are very helpful in discerning the source trajectories during the \wfirst~mission (see the first astrometry microlensing event in Figure \ref{exam1}), and fully characterizing the microlensing lightcurves with high confidence. In Figure \ref{exam2}, we show three more simulated astrometric microlensing events due to ISMBHs as detected by \wfirst,\ by assuming additional sparse observations as discussed above. In these plots the extra data points are depicted in green. We note that the astrometry data points during the time gap (green points) can jump to the observing seasons (shown by the red points) because of the added noise in the simulated data.

In the next section, we evaluate the expected errors in the physical parameters of ISMBHs detected through astrometric microlensing by the \wfirst~telescope. \\

\section{Observations of astrometric  microlensing}\label{stat}
To study detection and characterization of the ISMBHs by microlensing observations during the \wfirst~mission, we extend our simulation and make a big ensemble of detectable astrometric microlensing events. 

Since the mass function for ISMBHs are not well determined, so we consider several different mass functions. A simple form for ISMBHs' mass function is a uniform function versus mass in the range of $M_{\rm l} \in [2,~50] M_{\odot}$. Through modeling of black holes,  \citet{2022ApJSicilia} have found that the mass function of ISMBHs is almost flat up to $50 M_{\odot}$. Additionally, we examine three more mass functions, which are log-uniform ($dN/dM \propto 1/M$) and power-law ($dN/dM \propto M^{-0.5}$, and $dN/dM\propto M^{-2}$) ones.

Other parameters are determined according to their distribution functions, as explained in the previous papers \citep[see, e.g., ][]{sajadian_LSST,2017Marc}. {For each mass function, we perform the simulations two times, i.e., with and without considering sparse observations during the $\sim$2.3-year time gap. }

We choose the discernible events. Our criteria for detectability are (i) $\Delta \chi^{2} (=\big|\chi^{2}_{\rm{base}}-\chi^{2}_{\rm{real}}\big|) >800$ for photometry data points, and (ii) at least three photometry data points above the baseline by $4 \sigma_{\rm m}$, where $\sigma_{\rm m}$ is the photometric accuracy. In Figure \ref{hist}, we show the normalized (fractional) distributions for four observing parameters including $t_{\rm E},~m_{\rm{base}},~t_{0},~u_{0}$ of detectable microlensing events due to ISMBHs (by considering a uniform mass function and sparse observations during the large time gap) in green color. In order to study for which kind of these microlensing events the physical parameters of their lens objects are measurable with reasonable accuracy, we also plot the corresponding normalized distributions of events with the relative errors in the lens mass, distance, and proper motion $\leq 5\%$ (black stepped curves).

Accordingly, detectable microlensing events due to ISMBHs have the average timescale of $\left<t_{\rm E}\right>=303$ days and their average source magnitude at the baseline is $\left<m_{\rm{base}}\right>=20.1$ mag. Discerning these microlensing light curves (by adding extra observations during the large time gap) does not highly depend on the time of the closest approach and the lens impact parameter. The events with measurable physical parameters of their lens objects have on average smaller lens impact parameters (by $0.13$), and mostly happen during either three first or three last observing seasons of the \wfirst\ telescope.

For each discernible event, we determine the errors in the physical parameters of microlenses through calculating Fisher and Covariance matrices \citep[see, e.g., ][]{1996Boutreux,1999SalimGould,2015AJsajadian}. In this regard, we make several simple assumptions which are listed here. (i) We separate the photometry and astrometry measurements completely and calculate two Fisher matrices corresponding to these measurements, $\boldsymbol{\mathcal{A}}$, and $\boldsymbol{\mathcal{B}}$ for each event. (ii) We assume that the lensing parameters such as $t_{0}$, $u_{0}$, $t_{\rm E}$, and $\xi$ are determined through photometry observations well and their real values are used for astrometric modeling. In fact, the photometric accuracy is better than the astrometric accuracy. (iii) We ignore the parallax effect on the source trajectories, which are too small to be measured (compare the dotted and dashed blue lines in right panels in Figures \ref{exam1}, and \ref{exam2}). (iv) We ignore the finite source effects on both microlensing lightcurves and astrometric shifts in the source position. (v) We assume that the source distances from the observer, $D_{\rm s}$, are determined by other observations, and we do not need to tune them through microlensing observations. {For instance, the Gaia observations provide stellar parallax distances for some source stars.}

Photometry and astrometry Fisher matrices are: 
\begin{eqnarray}
\mathcal{A}_{ij}&=&\sum_{k=1}^{N} \frac{1}{\sigma_{\rm m}^{2}(t_{\rm k})} \frac{\partial^{2} m_{\rm s}(t_{\rm k})}{\partial p_{i}\partial p_{j}}, \nonumber \\
\mathcal{B}_{ij}&=&\sum_{k=1}^{N} \frac{1}{\sigma_{\rm a}^{2}(t_{\rm k})}\Big( \frac{\partial^{2} \theta_{\rm s, n1}(t_{\rm k})}{\partial q_{i}~\partial q_{j}} +  \frac{\partial^{2} \theta_{\rm s, n2}(t_{\rm k})}{\partial q_{i}~\partial q_{j}}\Big),
\end{eqnarray}

\noindent where, $m_{\rm s}(t_{\rm k})=m_{\rm{base}}-2.5 \log_{10}\big[f_{\rm b} A(t_{\rm k}) +1-f_{\rm b} \big]$ is the apparent source magnitude at the given time $t_{\rm k}$. $f_{\rm b}$ is the blending factor in $W149$ filter, $m_{\rm{base}}$ is the baseline magnitude without lensing effect in that filter (its distribution for detectable events is shown in the second panel of Figure \ref{hist}). $p_{i}$s, and $q_{i}$s are observable parameters that affect on photometry and astrometry measurements ($m_{\rm s},~\boldsymbol{\theta}_{\rm s}$), respectively. 

{\bf Observable parameters:} A microlensing light curve by considering the parallax effect can be modeled with $7$ parameters which are: $p_{i} \in t_{0},~u_{0},~t_{\rm E},~\xi,~f_{\rm b},~m_{\rm{base}},~\pi_{\rm E}$. The finite source effect can be ignored in long-duration microlensing events due to ISMBHs, so we put aside this effect while calculating $\boldsymbol{\mathcal{A}}$. The source apparent trajectory on the sky plane can be modeled with 3 parameters: $q_{i} \in \theta_{\rm E},~ \mu_{\rm s, n1},~\mu_{\rm s, n2}$. 

We calculate Fisher matrices numerically. Their inverses (i.e., covariance matrices, $\boldsymbol{\mathcal{A}}^{\bf -1}$ and $\boldsymbol{\mathcal{B}}^{\bf -1}$) are derived using the Python module \texttt{Numpy} \footnote{\url{https://numpy.org/}}. The square roots of diagonal elements are the errors in the observable parameters, e.g., $\sigma_{p_{i}}= \sqrt{\mathcal{A}^{\bf -1}_{~ii}}$ and $\sigma_{q_{i}}=\sqrt{\mathcal{B}^{\bf -1}_{~ii}}$, {and non-diagonal elements are the correlation coefficients between errors in the parameters.}

\noindent Taking these errors into account, we determine the errors in the physical parameters of ISMBHs, which is explained in the next subsection.\\  
\begin{figure*}
\includegraphics[width=0.49\textwidth]{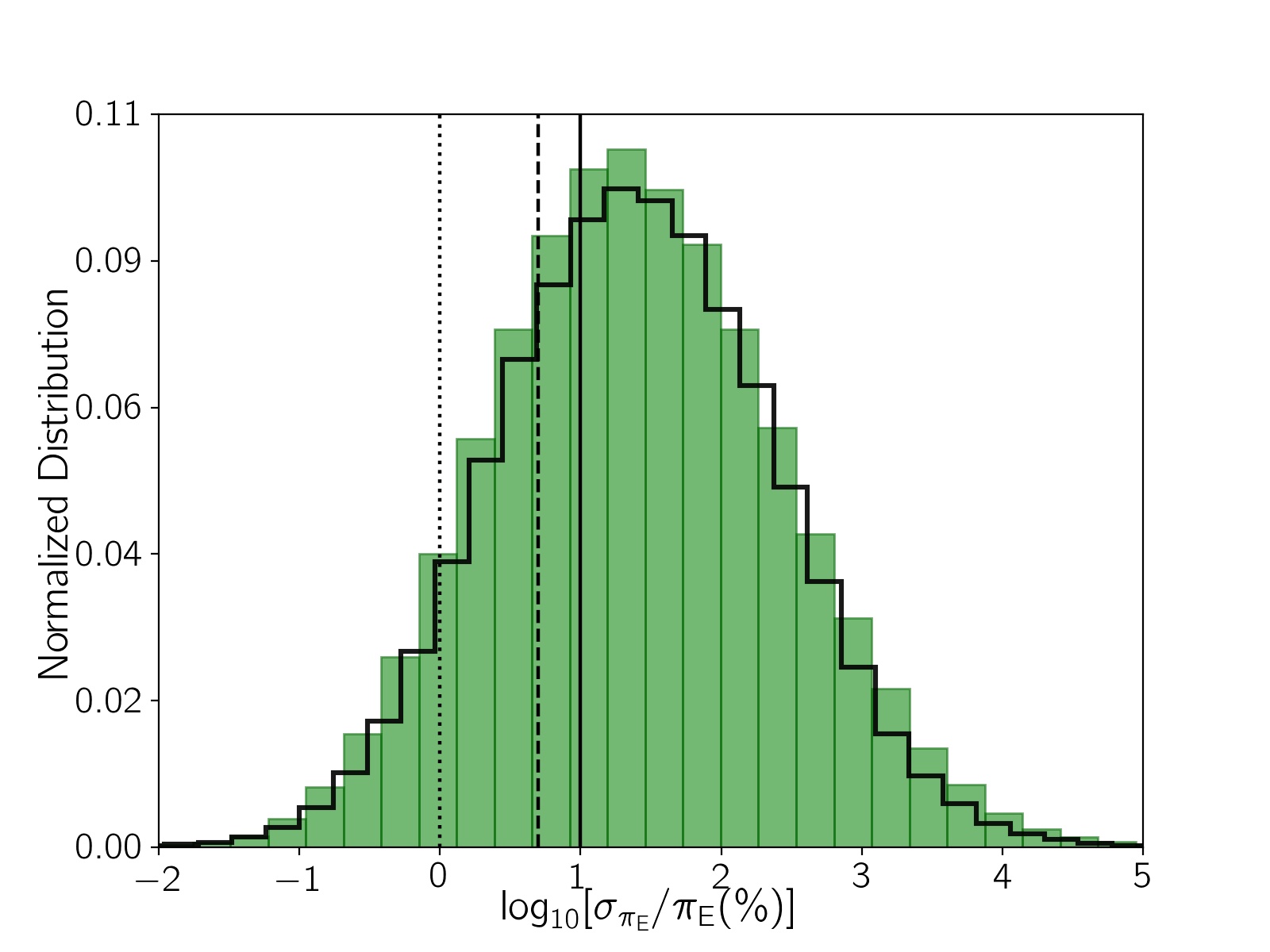}
\includegraphics[width=0.49\textwidth]{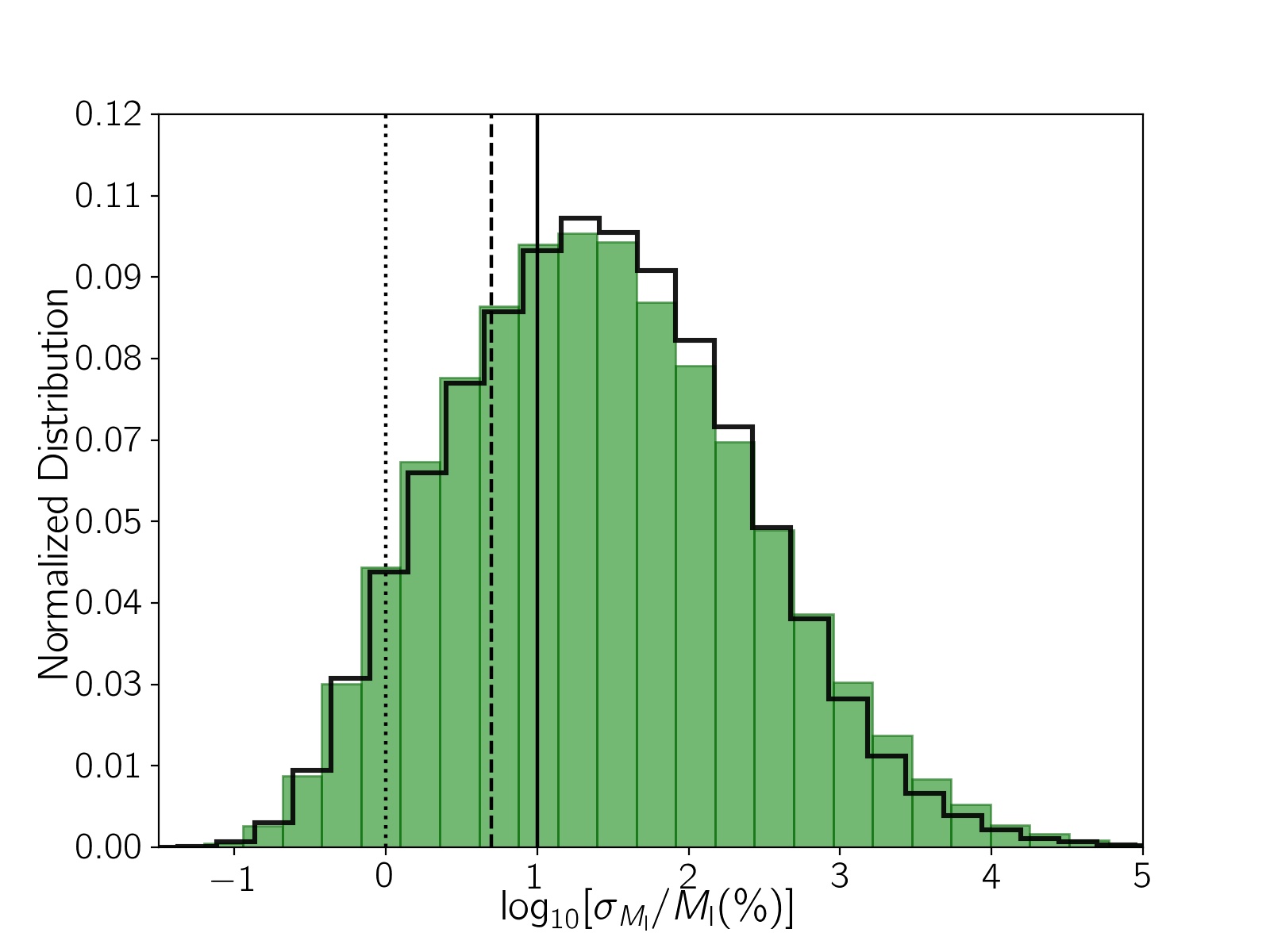}
\includegraphics[width=0.49\textwidth]{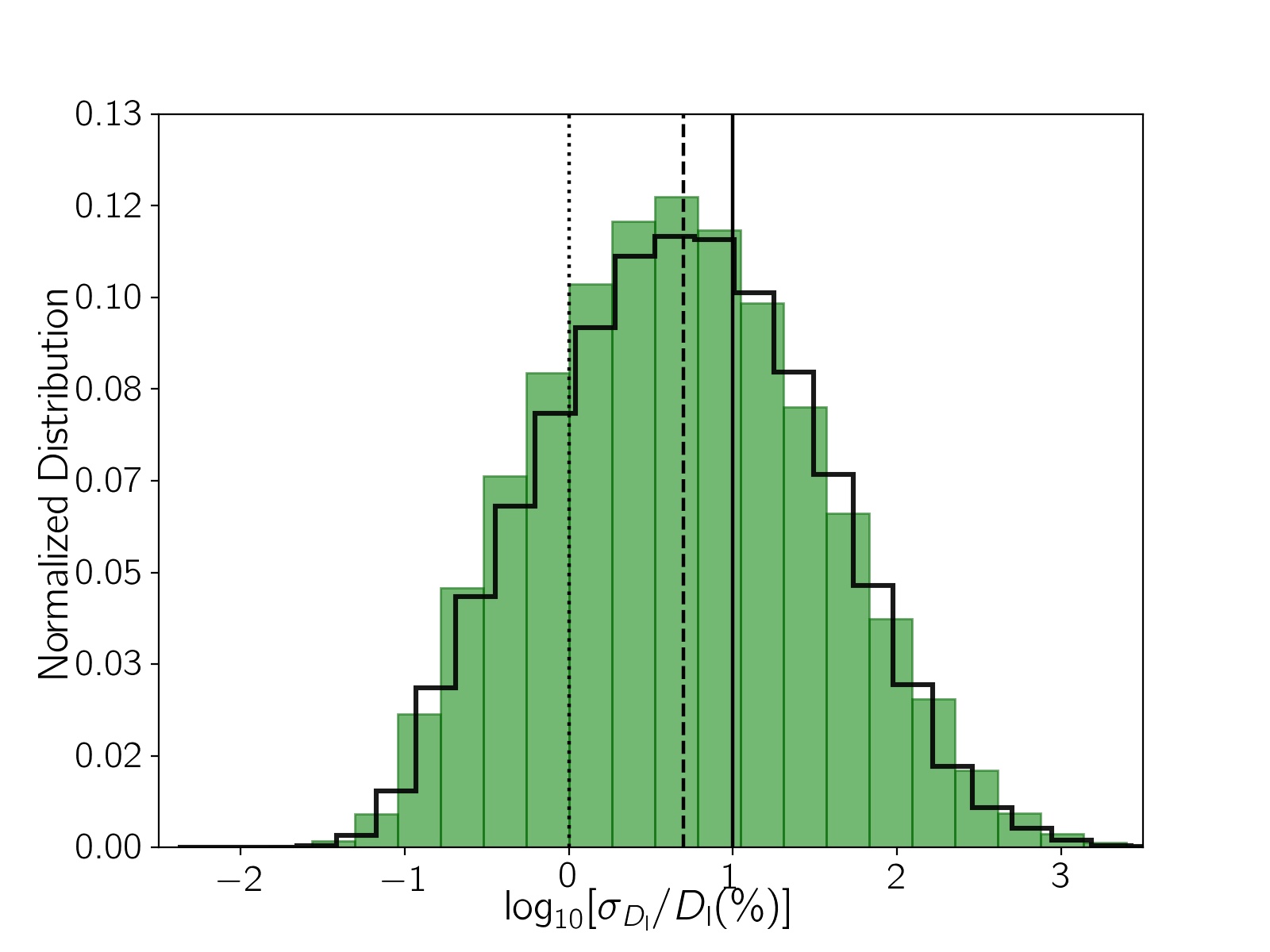}
\includegraphics[width=0.49\textwidth]{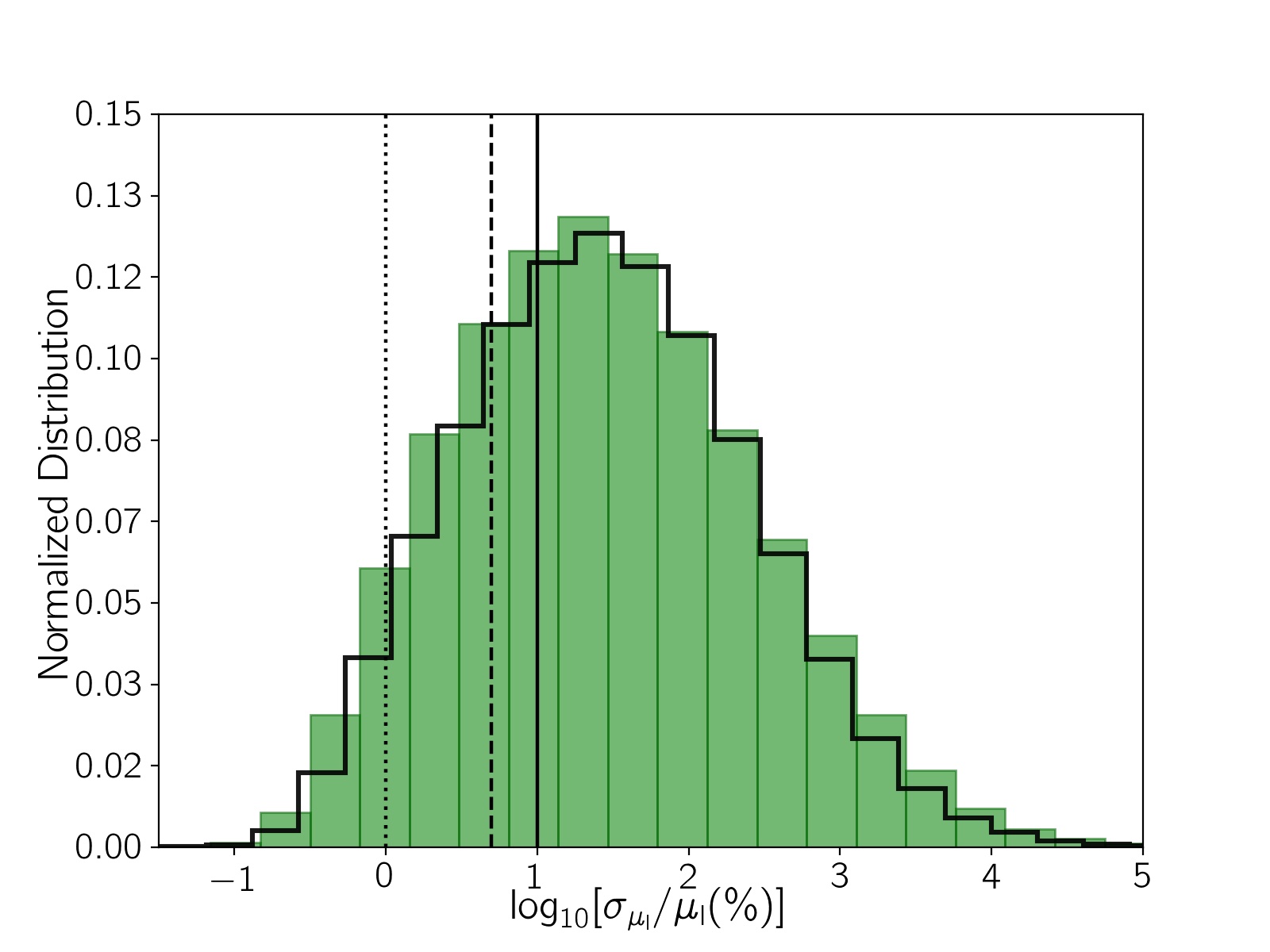}
\caption{The fractional distributions of the relative errors in the normalized parallax amplitude, the lens mass, the lens distance, and the lens proper motion for a big samples of microlensing events due to ISMBHs detectable by the \wfirst~telescope with (green distributions) and without (black step ones) considering sparse observations when the Galactic bulge is observable during the large time gap. The vertical (solid, dashed and dotted) lines show the thresholds of the relative errors $10\%$, $5\%$, and $1\%$, respectively. The samples due to both distributions have the same entrances. }
\label{histo}
\end{figure*}

\subsection{Errors in the physical parameters}
According to Equation \ref{tee}, the lens mass and its error as a function of observable parameters are:  

\begin{eqnarray}
M_{\rm l}&=&\frac{\theta_{\rm E}}{\kappa~ \pi_{\rm E}}, \nonumber\\
\sigma_{\rm M_{\rm l}}&=& M_{\rm l} \sqrt{\Big(\frac{\sigma_{\theta_{\rm E}}}{\theta_{\rm E}}\Big)^{2} + \Big(\frac{\sigma_{\pi_{\rm E}}}{\pi_{\rm E}}\Big)^{2}}, 
\end{eqnarray}

\noindent where $\sigma_{\rm M_{\rm l}}$, $\sigma_{\theta_{\rm E}}$, and $\sigma_{\pi_{\rm E}}$ are the error in the lens mass, error in the angular Einstein radius, and the error in normalized parallax amplitude, respectively. {We note that there is no correlation between $\sigma_{\pi_{\rm E}}$ and $\sigma_{\theta_{\rm E}}$, because these two parameters are determined from photometry and astrometry Fisher matrices independently.} The next parameter is the lens distance which is given by:
\begin{eqnarray}
\frac{1}{D_{\rm l}}&=&\frac{1}{D_{\rm s}}~+~\frac{\pi_{\rm E}~\theta_{\rm E}}{\rm{au}}, \nonumber \\
\sigma_{D_{\rm l}}&=&D_{\rm l}\frac{D_{\rm s}-D_{\rm l}}{D_{\rm s}}~\frac{\sigma_{\rm M_{\rm l}}}{M_{\rm l}}, 
\label{dll}
\end{eqnarray} 

\noindent Here, we assume that the error in source distance is very small and can be ignored. The last parameter is the lens angular velocity components which are: 
\begin{eqnarray}
\mu_{\rm l, n1}&=& \mu_{\rm s, n1}~-~\frac{\theta_{\rm E}}{t_{\rm E}}~\cos \xi,  \nonumber\\
\mu_{\rm l, n2}&=& \mu_{\rm s, n2}~-~\frac{\theta_{\rm E}}{t_{\rm E}}~\sin \xi, 
\end{eqnarray}
\begin{deluxetable*}{cccccccccc}
\tablecolumns{10}
\centering
\tablewidth{0.95\textwidth}\tabletypesize\footnotesize\tablecaption{Statistical information about simulated microlensing events due to ISMBHs detectable with the \wfirst~telescope {by assuming different ISMBHs mass functions.}}
\tablehead{\colhead{$~~~$}& \colhead{$\sigma_{t_{\rm E}}\big/t_{\rm E}$} & \colhead{$\sigma_{\pi_{\rm E}}\big/\pi_{\rm E}$} & \colhead{$\sigma_{\theta_{\rm E}}\big/\theta_{\rm E}$} & \colhead{$\sigma_{M_{\rm l}}\big/M_{\rm l}$} & \colhead{$\sigma_{D_{\rm l}}\big/D_{\rm l}$} & \colhead{$\sigma_{\mu_{\rm s}}\big/\mu_{\rm s}$} & \colhead{$\sigma_{\mu_{\rm l}}\big/\mu_{\rm l}$} &\colhead{$\epsilon_{\rm m}(\%)$} & \colhead{$N_{e,\rm{BHs}}$}} \\
\startdata \\
\multicolumn{10}{c}{${\bf dN/dM =\rm{const}}$}\\ \multicolumn{10}{c}{No~observations~during~the~time~gap}\\
$\leq1\%$&$23.60$ & $7.50$ & $85.56$ & $6.11$ & $21.15$ & $99.67$ & $5.16$ & $4.21$ & $2$\\
$\leq5\%$&$53.26$ & $24.35$ & $99.32$ & $24.08$ & $50.59$ & $99.98$ & $22.32$ & $19.37$&$11$\\
$\leq10\%$&$65.91$ & $34.86$ & $99.88$ & $34.77$ & $64.11$ & $100.00$ & $33.00$ & $29.29$ &$17$\\
\\ \multicolumn{10}{c}{Sparse~observations~during~the time gap}\\
$\leq1\%$&$30.81$ & $8.32$ & $83.15$ & $6.93$ & $22.99$ & $99.66$ & $6.10$ & $5.15$ & $4$\\
$\leq5\%$&$63.72$ & $25.66$ & $98.85$ & $25.40$ & $52.37$ & $99.98$ & $24.27$ & $21.48$& $17$\\
$\leq10\%$&$76.00$ & $36.14$ & $99.75$ & $36.05$ & $65.26$ & $99.99$ & $34.98$ & $31.54$& $24$\\

\hline \\ \multicolumn{10}{c}{${\bf dN/dM \propto M^{-0.5}}$}\\  \multicolumn{10}{c}{No~observations~during~the time gap}\\
$\leq1\%$ &$22.20$ & $7.52$ & $75.03$ & $5.34$ & $19.43$ & $99.68$ & $4.38$ & $3.64$ & $2$\\
$\leq5\%$ &$49.88$ & $22.52$ & $98.29$ & $21.98$ & $45.97$ & $99.99$ & $20.34$ & $17.57$ & $12$\\
$\leq10\%$ &$62.02$ & $31.84$ & $99.65$ & $31.66$ & $59.07$ & $99.99$ & $29.94$ & $26.30$ &$18$\\
\\ \multicolumn{10}{c}{Sparse~observations~during~the time gap}\\
$\leq1\%$ &$25.77$ & $7.70$ & $71.64$ & $5.65$ & $19.49$ & $99.66$ & $4.94$ & $4.22$ & $3$\\
$\leq5\%$ &$56.57$ & $22.29$ & $97.40$ & $21.82$ & $45.21$ & $99.98$ & $20.81$ & $18.25$& $15$\\
$\leq10\%$ &$69.18$ & $31.33$ & $99.32$ & $31.15$ & $57.54$ & $99.99$ & $30.05$ & $26.75$ & $22$\\

\hline\\ \multicolumn{10}{c}{${\bf dN/dM \propto M^{-1}}$}\\ \multicolumn{10}{c}{No~observations~during~the time gap}\\
$\leq1\%$ & $21.89$ & $7.52$ & $71.23$ & $5.11$ & $18.85$ & $99.67$ & $4.19$ & $3.51$ & $3$\\
$\leq5\%$ & $48.83$ & $22.00$ & $97.82$ & $21.34$ & $44.75$ & $99.98$ & $19.79$ & $17.00$ & $14$\\
$\leq10\%$& $61.02$ & $31.20$ & $99.56$ & $30.97$ & $57.68$ & $99.99$ & $29.15$ & $25.56$& $21$\\
\\ \multicolumn{10}{c}{~Sparse~observations~during~the time gap}\\
$\leq1\%$ & $24.48$ & $7.55$ & $67.89$ & $5.30$ & $18.56$ & $99.67$ & $4.56$ & $3.92$& $3$\\
$\leq5\%$ & $54.23$ & $21.38$ & $96.75$ & $20.81$ & $43.22$ & $99.99$ & $19.85$ & $17.33$ & $15$\\
$\leq10\%$& $66.95$ & $30.00$ & $99.17$ & $29.79$ & $55.42$ & $100.00$ & $28.79$ & $25.61$& $22$\\

\hline\\  \multicolumn{10}{c}{${\bf dN/dM \propto M^{-2}}$}\\ \multicolumn{10}{c}{No~observations~during~the time gap}\\
$\leq1\%$ &  $21.75$ & $7.15$ & $59.45$ & $4.51$ & $16.60$ & $99.69$ & $3.83$ & $3.34$ & $3$\\
$\leq5\%$ & $49.50$ & $19.65$ & $95.20$ & $18.83$ & $39.16$ & $99.99$ & $17.93$ & $15.53$ &$12$ \\
$\leq10\%$ & $62.21$ & $27.56$ & $98.69$ & $27.24$ & $50.89$ & $100.00$ & $26.30$ & $23.07$ & $18$\\
\\ \multicolumn{10}{c}{Sparse~observations~during~the time gap}\\
$\leq1\%$ & $21.00$ & $7.57$ & $62.54$ & $4.46$ & $17.91$ & $99.67$ & $3.71$ & $3.31$ &  $3$\\
$\leq5\%$ & $46.86$ & $21.33$ & $96.58$ & $20.35$ & $42.25$ & $99.98$ & $18.81$ & $16.08$ & $15$ \\
$\leq10\%$ & $58.57$ & $29.98$ & $99.28$ & $29.61$ & $54.68$ & $100.00$ & $27.93$ & $24.39$ & $23$\\
\tablecomments{Each entry represents the persentage of simulated events with the desired relativel error (specified in its row) be less than the given threshold (determined in its column). $\epsilon_{\rm m}$ is the~\wfirst~efficiency for measuing the lens mass, distance, and its proper motion with the relative errors less than the given threshold. {The last column reports the estimated number of ISMBHs that can be detected in the \wfirst\ observations by considering different mass functions, as explained in Subsection \ref{stats}.}}
\enddata
\label{tabfin} 
\end{deluxetable*}

\noindent Accordingly, the errors in the lens angular velocity components are given by:  
{
	\begin{eqnarray}
	\sigma_{\rm l, n1}^{2}=\sigma_{\rm s, n1}^2&+&\mu^{2}_{\rm rel,\odot}\cos^{2}\xi\Big[\big(\frac{\sigma_{\theta}}{\theta_{\rm E}}\big)^{2}+\big(\frac{\sigma_{t}}{t_{\rm E}}\big)^{2}\nonumber\\&+&\big(\frac{\sigma_{\xi}}{\cot \xi}\big)^{2}-2\frac{\sigma_{t}}{t_{\rm E}}\frac{\sigma_{\xi}}{\cot \xi}\hat{\mathcal{A}}^{\bf -1}_{~ij}  \Big],~~~~~\nonumber \\
	\sigma_{\rm l, n2}^{2}=\sigma_{\rm s, n2}^2&+&\mu^{2}_{\rm rel, \odot}\sin^{2}\xi\Big[\big(\frac{\sigma_{\theta}}{\theta_{\rm E}}\big)^{2}+\big(\frac{\sigma_{t}}{t_{\rm E}}\big)^{2}\nonumber\\&+&\big(\frac{\sigma_{\xi}}{\tan \xi}\big)^{2} -2\frac{\sigma_{t}}{t_{\rm E}}\frac{\sigma_{\xi}}{\tan \xi}\hat{\mathcal{A}}^{\bf -1}_{~ij} \Big].~~~~~
	\end{eqnarray}}

\noindent where, $\sigma_{\rm l, i},~\sigma_{\rm s, i}$ are the errors in $i$th component of the lens and source angular velocity projected on the sky plane, {and $\hat{\mathcal{A}}^{\bf -1}_{~ij}=\mathcal{A}^{\bf -1}_{~ij} /\sqrt{\mathcal{A}^{\bf -1}_{~ii} \mathcal{A}^{\bf -1}_{~jj}}$ is the correlation coefficient between errors in $t_{\rm E}$, and $\xi$.} The errors in the lens and source proper motion can be determined using the errors in their components.

\subsection{Results}\label{conclu}
{The normalized distributions for four relevant parameters (i.e., $t_{\rm E}$, $m_{\rm{base}}$,  $t_{0}$, and $u_{0}$) for simulated events whose relative errors in the lens mass, distance and proper motion are $\leq 5\%$, are shown in Figure \ref{hist} with black step lines. Accordingly, longer microlensing events from brighter source stars, whose times of the closest approach happen during either the first three or the last three observing seasons are more favourable for the measurement of the physical parameters of the lens objects with reasonable accuracy.}

In Figure \ref{histo}, we show the normalized distributions of the relative errors in the physical parameters of ISMBHs (as microlenses), resulting from Monte Carlo simulations, by considering a uniform mass function for ISMBHs. {Green and black distributions are related to detectable events by the \wfirst\ telescope with and without considering sparse data points during the time gap, respectively. }
These parameters are the normalized parallax amplitude, the lens mass, the lens distance and the lens proper motion. The threshold amounts of the relative errors in the given parameters of $10\%$, $5\%$, and $1\%$ are depicted with solid, dashed, and dotted lines.  
{Accordingly, adding extra observations during the time gap (one hour of observations every 10 days when the Galactic bulge is observable) improves the relative errors in all physical parameters, especially the lens distance from the observer.}\\

\noindent For numerical evaluation, in Table \ref{tabfin} we give the percentages of simulated detectable events with the relative errors (specified in the first row) less than the given thresholds (i.e., $1, 5, 10\%$ as mentioned in the first column) are reported. {Hence, sparse observations during the time gap improve the \wfirst\ efficiencies by $\sim1\%$, $\sim 2\%$, and $\sim 2\%$ for measuring the physical parameters by the relative errors less than $1,~5,~10\%$, respectively. } 

In $20$-$25\%$ detectable events, the lens mass can be determined with the relative error less than $5\%$. These events have smaller relative errors in the lens distance, because the factor $(D_{\rm s}-D_{\rm l})/D_{\rm s}$ is less than one. 

The source proper motion can be determined by monitoring the source positions during 6 observing seasons (with a $15$ min cadence) of the \wfirst~mission even without taking sparse data points during the $\sim$2.3-year time gap very well. Nevertheless, the lens proper motion can be determined with the relative error less than $5\%$ in $19$-$24\%$ of these events.\\

Even though ISMBHs produce long-duration microlensing events, which are suitable for discerning the annual parallax effects, the normalized parallax amplitude, $\pi_{\rm E}$, decreases with increasing the lens mass. Hence, the parallax effect can be discerned in these long-duration microlensing events with the relative errors less than $5\%$ only in $21$-$26\%$ of all detectable events.\\

\begin{figure*}
\includegraphics[width=0.49\textwidth]{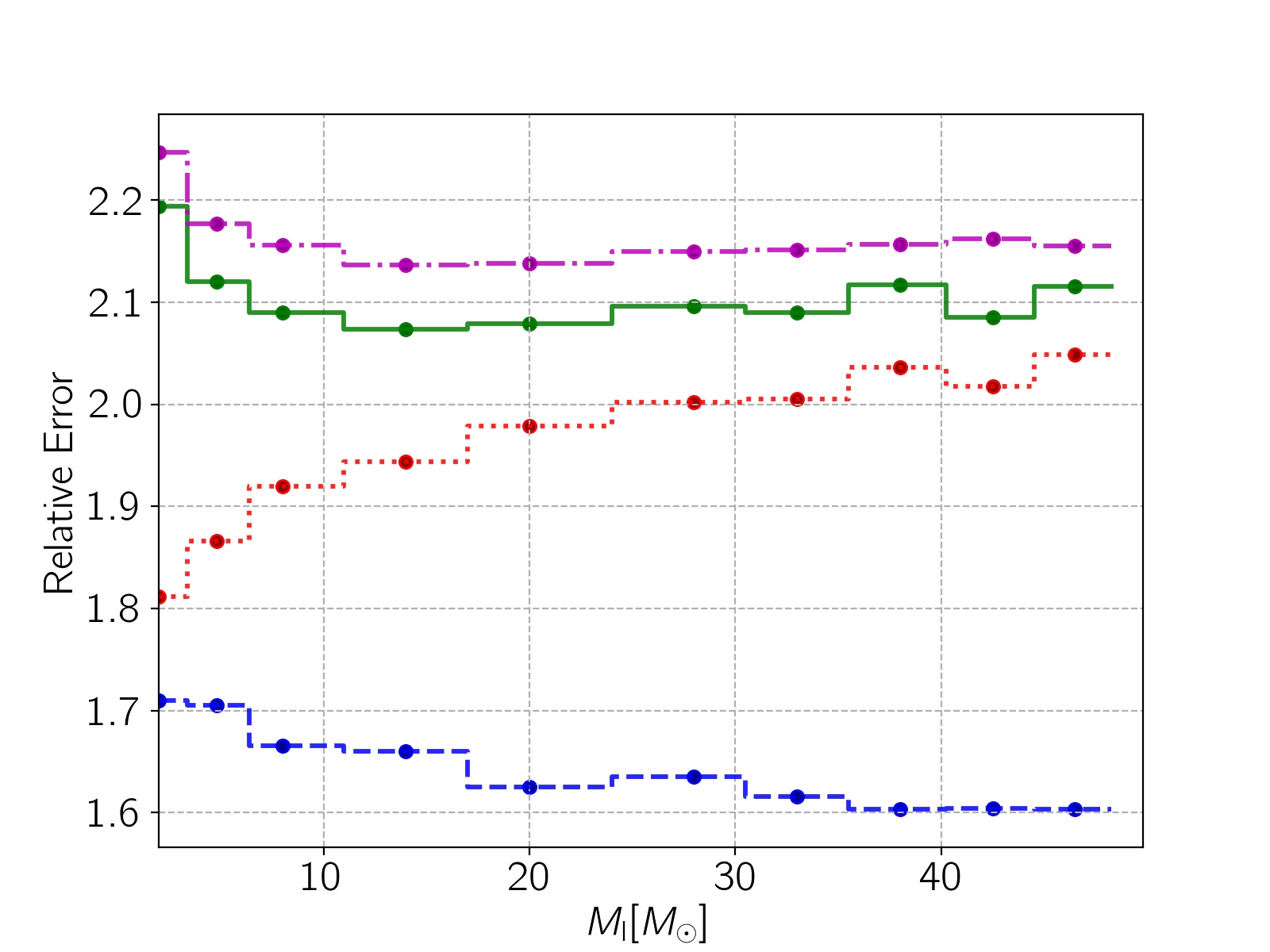}
\includegraphics[width=0.49\textwidth]{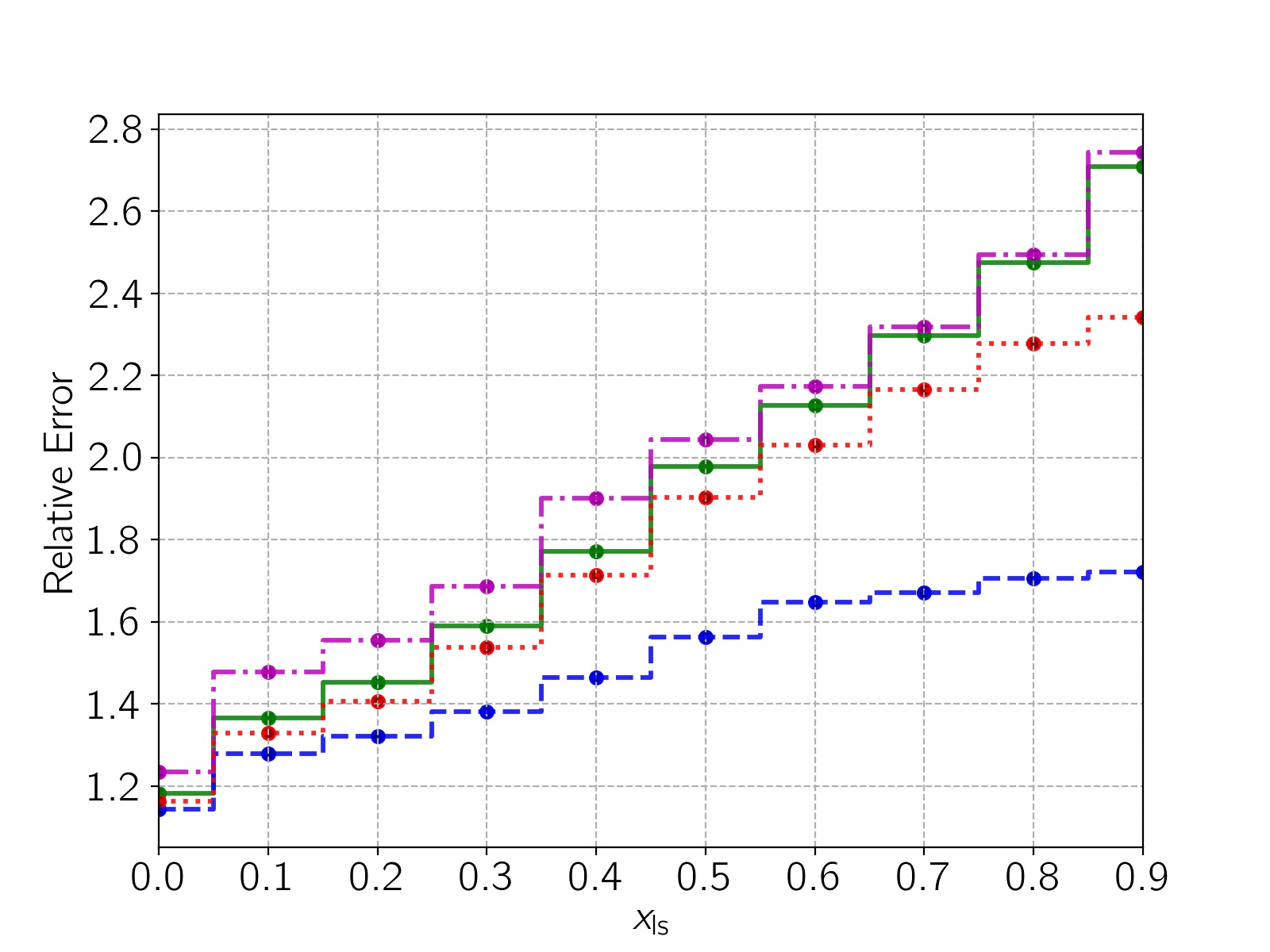}
\includegraphics[width=0.49\textwidth]{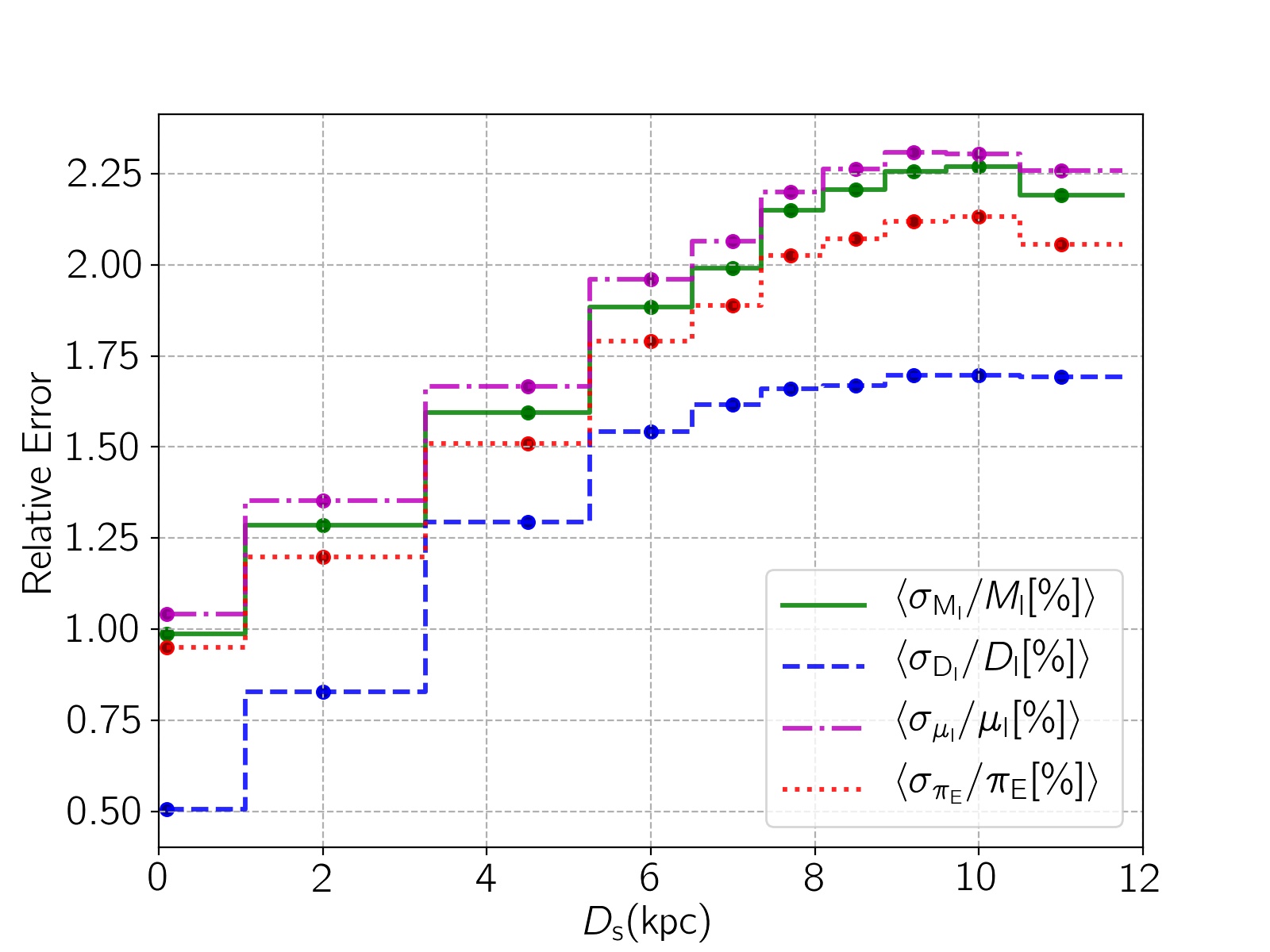}
\includegraphics[width=0.49\textwidth]{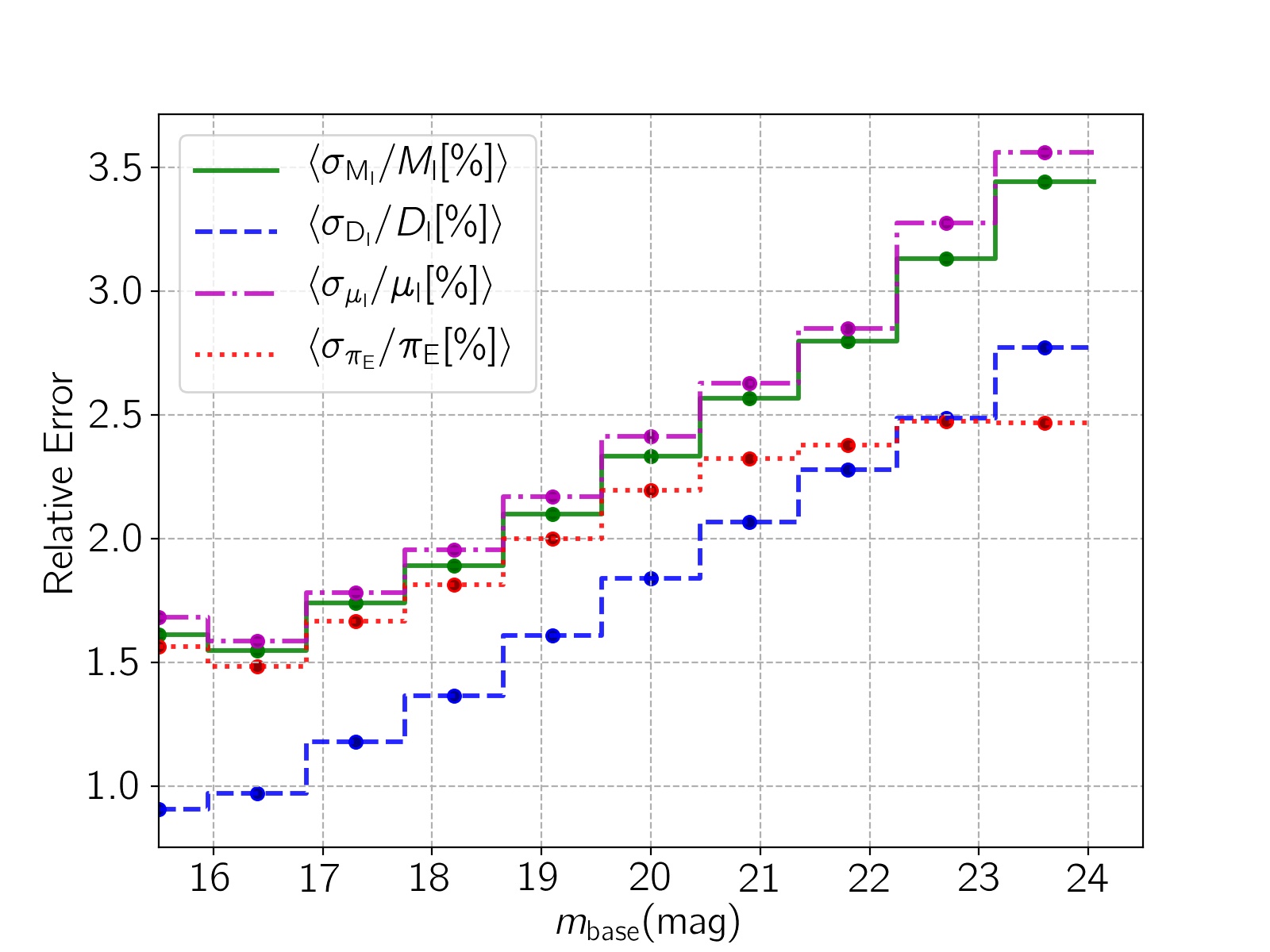}
\caption{The dependence of the average relative errors in the lens mass (solid green lines), the lens distance (dashed blue lines), its proper motion (dot-dashed magenta lines), and the normalized parallax amplitude (dotted red lines) versus the lens mass, the ratio of the lens distance to the source distance from the observer $(x_{\rm{ls}})$, the source distance, {and the source apparent magnitude at the baseline}.}
\label{error}
\end{figure*}

In order to determine which kinds of ISMBHs might be well characterized through astrometric microlensing observations with the \wfirst~telescope, we show the dependence of the relative errors in the lens mass, the lens distance, its proper motion, and the parallax amplitude to $M_{\rm l}$, $x_{\rm{ls}}$, $D_{\rm s}$, and $m_{\rm{base}}$ in Figure \ref{error}, in different panels, respectively. For these plots, we only use the events with the relative errors less than $5\%$. There are several factors which determine their dependencies.

\noindent According to the first panel, the relative error in the lens mass minimize when $M_{\rm l} \simeq 10$-$25 M_{\odot}$. Increasing the lens mass has two against effects: (i) The lens mass enhances the Einstein crossing time and decreases the average photometry errors. Because more data points are taken while the source is being lensed, and less data points are recorded over the baseline. (ii) Enhancing the lens mass decreases the normalized parallax amplitude $\pi_{\rm E}$ significantly, and makes hard measure it (see the dotted red step line in the top panel). This point was also expressed by \citet{2020Karolinski} and while modeling OGLE-2006-BLG-044 microlensing event. For that reason, the optimum value for the lens mass with least errors is neither the least (2-3 solar mass), nor the most (40-50 solar mass). The relative error in the lens distance decreases with the lens mass. In fact, by increasing the lens mass $x_{\rm{ls}}$ enhances to keep the Einstein crossing times close to reasonable values for detection.  

The relative error in the lens proper motion weakly depends on the lens mass. In fact, $\sigma_{t_{\rm E}}/t_{\rm E}$ is an increasing function versus the lens mass. By fixing the observing time and cadence (considering a determined observing platform) and increasing $t_{\rm E}$, its error increases. In total, the relative errors in the lens physical parameters enhance with the lens mass slowly.  

The second panel of Figure \ref{error} shows the relative errors in the lens mass, lens distance, its proper motion, and the parallax amplitude versus $x_{\rm{ls}}=D_{\rm l}/D_{\rm s}$. The smaller $x_{\rm{ls}}$ make larger $\pi_{\rm E}$ and $\theta_{\rm E}$, with smaller observing errors. That increases the relative error in the lens mass versus $x_{\rm{ls}}$. However, this enhancement is slower in the relative error in the lens distance, because of the factor $(D_{\rm s}-D_{\rm l})/D_{\rm s}$ in Equation \ref{dll}. 

In the next panel of Figure \ref{error}, we show the dependence of the relative errors with the source distance from the observer. The source distance decreases $\pi_{\rm E}$, and $\theta_{\rm E}$, which increases the relative errors in the lens mass and its distance. We note that decreasing the parallax amplitude increases both errors in the parallax amplitude, and $\xi$. 
Comparing these panels, we find that the effect of the source distance and the lens relative position $(x_{\rm ls})$ on the errors is higher than the effect of the lens mass. 

{In the last panel, the relative errors versus the apparent magnitude of the source star at the baseline are depicted. As shown here, they enhance with the source magnitude. Both \wfirst\ photometric and astrometric errors increase with the apparent magnitude of source stars. Worse accuracies cause higher relative errors in the lens physical parameters.}

Therefore, long-duration microlensing events due to ISMBHs with the mass $M_{\rm l} \simeq 10$-$25 M_{\odot}$, close to the observer ($x_{\rm ls}\lesssim 0.5$) while the source is inside the Galactic disk $(D_{\rm s}\lesssim 6 \rm{kpc})$ can be characterized with the least errors.

{\subsection{Different mass function for ISMBHs}}
{We know that there is no accurate mass function for ISMBHs based on observations yet, so we perform the simulation by considering several mass functions for ISMBHs, which are given in the following:  
\begin{eqnarray}
\frac{dN}{dM} &=& \rm{const.}, \nonumber\\
\frac{dN}{dM} &\propto& 1\big/\sqrt{M},\nonumber\\
\frac{dN}{dM} &\propto& M^{-1},  \nonumber\\
\frac{dN}{dM} &\propto& M^{-2}. 
\label{massfunc}
\end{eqnarray}	
The results from simulations based on each of these mass functions are reported in Table \ref{tabfin}. Accordingly, by changing ISMBHs mass function, the \wfirst\ efficiency to measure the lens physical parameters can change up to $2$-$7\%$. Also, the first mass function makes more ISMBHs with mass $M_{\rm l} \in [10,~25] M_{\odot}$ than other mass functions. So it has larger efficiencies to measure the physical parameters of lens objects than others.}

 In the next subsection, we do some statistical estimations about detecting and characterizing such events during the \wfirst~mission.

\subsection{Statistical estimations}
\label{stats}
The number of microlensing events that the \wfirst~telescope will detect is $N_{\rm e, tot}=27000$, which were estimated in \citet{2019ApJSwfirst,2020Johnson}. Here, we want to evaluate what fraction of this total number of microlensing events detectable by the \wfirst~telescope are due to ISMBHs. In this regard, there are two factors: (i) the optical depth, and (ii) the average microlensing duration which are discussed in the following.  

\noindent (i) The number of detectable microlensing events is proportional to the optical depth. The microlensing optical depth at a given line of sight $(l,~b)$ and one specified distance from the observer, $(D)$, is proportional to the lens mass $M_{\rm l}$, because it is given by:
\begin{eqnarray}
\frac{d\tau(l,~b,~D)}{dD} =\pi~\theta_{\rm E}^{2}~n(l,~b,~D)~D^{2},
\end{eqnarray}
where, $(l,~b)$ are the Galactic longitude and latitude, respectively. $n(l,~b,~D)$ is the number density of stars in our galaxy which is the Galactic mass density divided by the average stellar mass.   

\noindent Accordingly, the ratio of the optical depth (and as a result the number of microlensing events) due to ISMBHs to the overall optical depth due to all potential lens objects can be estimated by:  
\begin{eqnarray}
\mathcal{F}_{1}=\int_{20 M_{\odot}}^{\infty}M_{\rm l}~\eta(M_{\rm l})~dM_{\rm l} \Big/ \int_{13M_{\rm J}}^{\infty}M_{\rm l}~\eta(M_{\rm l})~dM_{\rm l},  
\label{factor1}
\end{eqnarray}
\noindent where,$M_{\rm J}$ is the Jupiter mass, $\eta(M_{\rm l})$ is the initial mass function in the Galactic disk. In fact, $\mathcal{F}_{1}$ determines the contribution of the ISMBHs in producing the effective lensing surface in comparison with the total lensing surfaces covered by all possible Einstein rings. {In Equation \ref{factor1}, we use the fact that stars with the initial mass $M > 20 M_{\odot}$ will convert to black holes. We ignore the contribution of black holes generated from primordial fluctuations in the early universe.}

In order to estimate $\mathcal{F}_{1}$, we take the initial mass function from the Besan\c{c}on model \citep{Robin2003,Robin2012}, and assume that all lens objects are inside the Galactic disk. This mass function is $\eta(M_{\rm l}) \propto M_{\rm l}^{-1.6}$ for $0.08\leq M_{\rm l}(M_{\odot}) \leq 1$, and $\eta(M_{\rm l}) \propto M_{\rm l}^{-3}$ for $M_{\rm l}(M_{\odot}) \geq 1$. The stars with $M_{\rm l}>20 M_{\odot}$ are converted to ISMBHs. For $13 M_{\rm J}<M_{\rm l}<0.08 M_{\odot}$ we take the Brown dwarf mass function, i.e., $M_{\rm l}^{-0.7}$ \citep{SONYCII, Luhman2004}. We do not include free floating planets, because of their negligible contribution. The upper limit should in reality be the mass due to the most massive star in the Galactic disk. We set this upper limit to infinity, because the mass function for $M>1 M_{\odot}$ decreases as $M^{-3}$, so it tends to zero fast. Accordingly, we find $\mathcal{F}_{1} = 0.019$.\\

(ii) The microlensing event rate is proportional to $\left< \epsilon(t_{\rm E})\big/t_{\rm E}\right>$, which specifies the inverse of the average duration of microlensing events. Here, $\epsilon(t_{\rm E})$ is the ~\wfirst~efficiency for detecting a microlensing event with the specified time scale $t_{\rm E}$, and was kindly provided by M. Penny. Since ISMBHs make longer microlensing events than usual events, we expect this factor for ISMBHs to be smaller than that due to all detectable microlensing events due to all potential lens objects. We define another factor: 
\begin{eqnarray}
\mathcal{F}_{2}= \left<\frac{\epsilon(t_{\rm E})}{t_{\rm E}}\right>_{\rm{BHs}} \Big/ \left<\frac{\epsilon(t_{\rm E})}{t_{\rm E}}\right>_{\rm{Total}}.
\end{eqnarray}
\noindent To estimate this factor, we simulate the microlensing events detectable by the \wfirst\ telescope, {and by adopting a uniform mass function for ISMBHs. However, we tune the ratio of the number of ISMBHs to the number of total objects  $\simeq0.0001$, as expected. In the simulation, the lens objects can be brown dwarfs, main-sequence stars and ISMBHs, and we obtain $\mathcal{F}_{2}=0.15,~0.11$ with and without considering sparse observations during the time gap, respectively. We note that considering extra observations enables us to detect ISMBHs in shorter microlensing events (the average $t_{\rm E}$ changes from $329$ days to $303$ days).}

Therefore, the \wfirst~telescope roughly will detect $N_{\rm e, BHs}= N_{\rm e, tot} \times \mathcal{F}_1 \times \mathcal{F}_2 \simeq 56$-$77$ microlensing events due to ISMBHs ({under the assumption that their masses are uniformly distributed in the range of $[2,~50] M_{\odot}$, and their contribution with respect to all lens objects is $0.0001$}). In 2-4, 11-17, and 17-24 of these events the physical parameters of ISMBHs (including their mass, distance and proper motion) can be determined with the relative errors less than $1\%$, $5\%$, and $10\%$, respectively, as reported in the last column of Table \ref{tabfin}.\\
  
For other mass functions, i.e., $dN/dM \propto M^{-\alpha}$ with $\alpha=0.5,~1,~2$, we get $\mathcal{F}_2=0.16$-$0.13,~0.17$-$0.16,~0.18$-$0.0.15$ (with and without adding extra observations during the time gap), respectively. The corresponding number of ISMBHs that can be detected and characterized through the \wfirst\ observations are reported in Table \ref{tabfin}.

\section{Conclusions} \label{conclusion}
In this work, we studied detection and characterization of ISMBHs through astrometric microlensing to be done by the upcoming microlensing survey by the \wfirst~telescope. 

This telescope has been planned to detect mostly short-duration microlensing events due to exoplanets beyond the snow line of main-sequence stars and free-floating exoplanets. 

Nevertheless, the duration of its mission is long enough to detect and characterize long-duration microlensing events, and its astrometric accuracy is high enough to discern the astrometric trajectories (and the dimensional lensing-induced shifts) of source stars. 

Here, we have done a comprehensive simulation of astrometric microlensing events due to ISMBHs that can be discerned by the \wfirst~telescope. For each simulated event we have calculated Fisher and Covariance matrices for photometry and astrometry measurements separately, and estimated the errors in observable parameters, and physical parameters of ISMBHs as well.  

Since the long time gap between \wfirst's first three observing seasons and the other three seasons would limit its efficiency and robustness for discerning and characterizing ISMBHs, we considered a small amount of additional observations when the Galactic bulge is visible during this time gap, by adding one hour of observations (4 data points) every 10 days when the Galactic bulge is detectable in our simulations. 
These additional observations amount to a total of about one day of observations with \wfirst.
We found that this small amount of extra observations increases \wfirst's efficiency of detecting and characterizing ISMBHs by $\sim 1-2\%$, and, more importantly,  improve the robustness of the results and help avoiding degenerate solutions. 

We note that photometric follow-up of these microlensing events with ground-based telescopes such as the Rubin Observatory during the time gap should also be helpful.The ground-based images may suffer from blending, but the higher-resolution images of \wfirst\ should help in correctly estimating the blending factor, thus providing useful data  for better characterization of the microlensing light curves. 

For long-duration microlensing events due to ISMBHs, the efficiency of \wfirst\ microlensing survey for measuring the physical parameters of the lens by considering different ISMBHs mass functions are summarized in Table \ref{tabfin}.

The efficiencies for measuring with better than $5\%$ uncertainty the lens mass, its distance, and its proper motion are $20$-$25\%$, $42$-$52\%$, and $19$-$24\%$, respectively, and the efficiency of measuring all the three parameters with better than $5\%$ uncertainty is $16$-$21\%$. 

ISMBHs produce long-duration microlensing events which are appropriate for discerning the annual parallax. On the other hand, the normalized parallax amplitudes decrease with $1/\sqrt{M_{\rm l}}$. Therefore, $\pi_{\rm E}$ can be measured with the relative error less than $5\%$ in only $21$-$26\%$ of these long-duration events.   

The relative errors in the physical parameters of ISMBHs increases with the source distance and $x_{\rm{ls}}=D_{\rm l}/D_{\rm s}$. The dependence of these relative errors to the lens mass is relatively weak and by changing the lens mass from $2$ to $50$ solar mass, these error changes less than $1\%$. On the whole, the least relative errors in the lens mass and its distance occurs when $M_{\rm l}\simeq 10$-$25 M_{\odot}$, $x_{\rm ls}\lesssim 0.5$, and $D_{\rm s}\lesssim 6$ kpc.  

We also statistically estimated the total number of microlensing events due to ISMBHs that can be detected and characterized with the \wfirst~telescope. By assuming different mass functions for ISMBHs (given in Equation \ref{massfunc}) in the range of $[2,~50] M_{\odot}$, we concluded that this telescope will detect $56$-$77$ long-duration microlensing events due to ISMBHs during its mission. Additionally, it can measure the physical parameters of ISMBHs with the relative errors less than $1\%$, $5\%$, and $10\%$ in 3-4, 15-17, 22-24 of these events, respectively. \\ 

{All simulations that have been done for this paper are available at:  \url{https://github.com/SSajadian54/AstrometryMicrolensing}}

\acknowledgments
Research efforts of KCS were supported by NASA through grants from STScI, under proposal IDs 14783, 15318 and 16200. {We thank the anonymous referee for his/her careful and useful comments, which improved the quality of the paper.}

\appendix
\begin{figure}
	\includegraphics[width=0.49\textwidth]{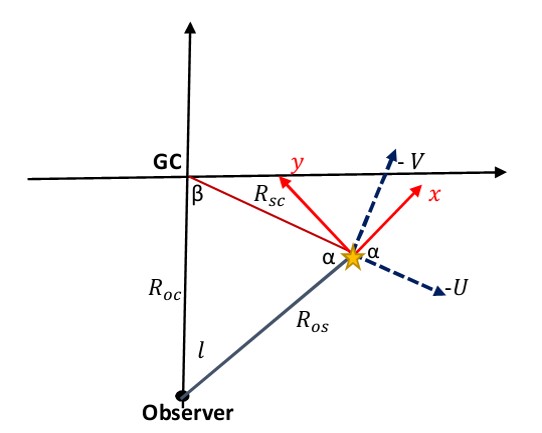}
	\caption{Figure shows the Galactic plane and two coordinate systems which are needed to project stellar velocities on the sky plane.}
	\label{fig1t}
\end{figure}
\section{Transforming coordinate systems} \label{append1}
In this section, we will review how to transform the stellar velocity from the Galactic coordinate frame to the observer one and project them on the sky plane.  

{In this Figure, the horizontal and vertical black lines describe the Galactic plane and make a right-hand coordinate system. We note that in this Figure the scales are not respected.} 

\noindent We consider a star in our galaxy with the galactic coordinate $(l,~b)$, i.e., the galactic longitude and latitude, respectively. Three points of the Galactic center (GC), the star position projected on the Galactic plane (yellow star) and the observer position (black filled point) make a triangle with the angles $l,~\alpha,~\beta$, as shown in Figure \ref{fig1t}. The length scales: $R_{\rm{oc}}$ the observer distance from the Galactic center, $R_{\rm{os}}$ the distance between the star position projected on the Galactic plane and the observer, and $R_{\rm{sc}}$ which is the distance between the Galactic center and the projected stellar position on the Galactic plane. $R_{\rm{sc}}$ can be given by: 
\begin{eqnarray}
R_{\rm sc} = \sqrt{R_{\rm oc}^{2} + R_{\rm os}^{2}- 2 R_{\rm os} R_{\rm oc} \cos(l)}.
\end{eqnarray}

\noindent where, $R_{\rm{os}}=D_{\star} \cos(b)$, and $D_{\star}$ is the star distance from the observer. Using the sinuous law in a triangle, we can derive the angle of $\beta$, as: 
\begin{eqnarray}
\sin (\beta)= \frac{R_{\rm os}}{R_{\rm sc}} \sin (l).
\end{eqnarray}
By having the Galactic longitude, we will calculate the angle of $\alpha$ as $\alpha= \pi-l-\beta$.\\ 

In simulations, we determine the stellar velocities in the Galactic coordinate, i.e., $(v_{\rm U}, v_{\rm V}, v_{\rm W})$, which are toward the Galactic center, in the direction of the Galactic rotation, and toward the Galactic north, respectively. These velocities include the global rotational velocity which is a function of the stellar distance from the Galactic center \citep[see, e.g.,][]{2009Rahal}, and velocity dispersion components which are functions of the stellar age, weakly mass, and the Galactic latitude \citep{1985ApJCarlberg_AVR,2020_1,2021sajadian_MVR}.     

\indent In the lensing formalism, we need the projected components of stellar velocities on the sky plane.  So we introduce another coordinate frame, $(x,~y,~z)$, which $z$-axis is parallel with $W$ (toward the Galactic north), and $(x,~y)$ describes the Galactic plane, as shown in Figure \ref{fig1t} with red vectors. We can easily convert the velocity components from Galactic coordinate frame to this new coordinate system, $(x, y, z)$, as:  

\begin{eqnarray}
v_{\rm x}&=& -\cos(\alpha)~v_{\rm U} -\sin(\alpha)~v_{\rm V},\nonumber \\
v_{\rm y}&=& +\sin(\alpha)~v_{\rm U} -\cos(\alpha)~v_{\rm V},\nonumber \\
v_{\rm z}&=& v_{\rm W},
\end{eqnarray}

Note that stars are not in the Galactic disk and their line of sight (los) with respect to the Galactic plane make the angle $b$, the Galactic latitude. So, we should apply another rotation around $y$-axis with $-b$ angle to obtain the components of stellar velocities projected on the sky plane normal to the line of sight toward the stellar position as:  
\begin{eqnarray}
v_{\rm{los}}&=& \cos(b)~v_{\rm x} +\sin(b)~v_{\rm z}, \nonumber\\
v_{\rm{n1}} &=& v_{\rm y},  \nonumber \\
v_{\rm{n2}} &=& -\sin(b)~v_{\rm x} + \cos(b)~v_{\rm z},
\end{eqnarray}

\noindent $n1$ and $n2$ are two unit vectors describe the sky plane. For projection of the Sun velocity, $\alpha_{\odot} \simeq \pi- l$, since $\beta_{\odot}\simeq 0$. For the observer orbit around the Sun, we easily consider a circular orbit with the radius of the astronomical unit.\\

\bibliographystyle{aasjournal}
\bibliography{paperBH}{}
\end{document}